\documentclass{emulateapj}
\usepackage{apjfonts}

\shorttitle{Study of RX~J1713.7$-$3946 with {\it Suzaku}}
\shortauthors{Tanaka et al.}

\begin{document}

\title{Study of Nonthermal Emission from SNR RX~J1713.7$-$3946 with {\it Suzaku}}

\author{Takaaki Tanaka\altaffilmark{1,2}, 
Yasunobu Uchiyama\altaffilmark{1}, 
Felix A. Aharonian\altaffilmark{3,4}, 
Tadayuki Takahashi\altaffilmark{1,5}, 
Aya Bamba\altaffilmark{1},
Junko S. Hiraga\altaffilmark{6}, 
Jun Kataoka\altaffilmark{7}, 
Tetsuichi Kishishita\altaffilmark{1,5},
Motohide Kokubun\altaffilmark{1}, 
Koji Mori\altaffilmark{8}, 
Kazuhiro Nakazawa\altaffilmark{5}, 
Robert Petre\altaffilmark{9}, 
Hiroyasu Tajima\altaffilmark{2}, 
and Shin Watanabe\altaffilmark{1}
}

\altaffiltext{1}{Department of High Energy Astrophysics, 
Institute of Space and Astronautical Science (ISAS), 
Japan Aerospace Exploration Agency (JAXA), 
3-1-1 Yoshinodai, Sagamihara, Kanagawa 229-8510, Japan}
\altaffiltext{2}{Kavli Institute for Cosmology and Particle Astrophysics, Stanford Linear Accelerator Center, 
2575 Sand Hill Road M/S 29, Menlo Park, CA 94025}
\altaffiltext{3}{Dublin Institute for Advanced Studies, 5 Merrion Square, Dublin 2, Ireland}
\altaffiltext{4}{Max-Planck-Institut f\"ur Kernphysik, PO Box 103980, 69029 Germany}
\altaffiltext{5}{Department of Physics, University of Tokyo, 7-3-1 Hongo, Bunkyo, Tokyo 113-0033, Japan}
\altaffiltext{6}{Cosmic Radiation Laboratory, RIKEN, 2-1 Hirosawa, Wako, Saitama 351-0198, Japan}
\altaffiltext{7}{Department of Physics, Tokyo Institute of Technology, 
Ohokayama, Meguro, Tokyo 152-8551, Japan}
\altaffiltext{8}{Department of Applied Physics, Universityof Miyazaki, 
1-1 Gakuen Kibana-dai Nishi, Miyazaki 889-2192, Japan}
\altaffiltext{9}{X-ray Astrophysics Laboratory, NASA Goddard Space Flight Center, 
Greenbelt, MD 20771}

\begin{abstract}
We present results obtained from a series of observations of the supernova remnant 
RX~J1713.7$-$3946 by the {\it Suzaku} satellite  which  cover about two-thirds of the
remnant surface.  Hard X-rays have been detected from each pointing up to $\sim 40~{\rm keV}$. 
The hard X-ray spectra are described by power-law functions with photon 
indices of $\sim 3.0$, which are larger than those in the energy region below 10~keV. 
The combination of the spatially-integrated XIS and HXD spectra 
clearly reveals a spectral cutoff in the  X-ray spectrum 
which  is linked to the maximum energy of accelerated  
electrons  emitting synchrotron radiation. 
The broad-band coverage of {\it Suzaku} 
observations  from 0.4 keV to 40 keV   allows  us to derive,  for the first time, 
the energy spectrum of parent  electrons in the cutoff region.  
The inferred cutoff energy in the spatially-integrated X-ray spectrum indicates that 
the electron acceleration in the remnant proceeds  close to the  Bohm-diffusion limit. 
We discuss  implications of the spectral and morphological properties of 
{\it Suzaku} data in the context of the origin of   nonthermal
emission.  The {\it Suzaku}  X-ray and the H.E.S.S.  TeV gamma-ray  data together  
hardly can be  explained within a  pure leptonic scenario, unless we introduce an 
additional  component of relativistic  electrons with softer energy spectrum. 
Moreover, the leptonic models  require  very weak magnetic field  which does not agree 
with the recently discovered filamentary structure and  short-term variability  features 
of the X-ray emitting region. 
The  hadronic  models with strong magnetic field provide perfect fits to the  observed 
X-ray and TeV gamma-ray spectra through the synchrotron radiation  
of electrons and {\it p-p} interactions of  protons,   but require 
special  arrangements of model parameters  to explain the lack of thermal  component
of X-ray emission.  For the morphology studies, we compare the X-ray an TeV gama-ray 
surface brightness maps using the {\it Suzaku} XIS and 
the H.E.S.S.  data. We confirm  the previously reported 
strong correlation between X-ray and  TeV gamma-ray emission  components. At the same time 
the {\it Suzaku}  data  reveal a deviation  from the general tendency,  namely,   the X-ray emission in 
the western rim regions  appears  brighter  than expected  from 
the average X-ray to gamma-ray  ratio. 
\end{abstract}


\section{INTRODUCTION}
Supernova remnants (SNRs) have long been considered to be likely acceleration sites of 
cosmic-ray particles below the energy of the {\it knee}, $\sim 10^{15}~{\rm eV}$. 
The energy supply to explain the energy density of comic rays is satisfied if $\sim 1$--10\% 
of the energy of each supernova is transferred to accelerated particles. 
Also, the well developed theory of diffusive shock acceleration nicely 
explains the universal power-law spectrum of cosmic rays (e.g. \citealt{BE87,malkov01}). 
Although synchrotron emission detected in the radio band supports this idea observationally, 
no evidence of acceleration to TeV energy had been observed until 
recently. 
During the last decade, such evidence was revealed through observations of 
X-rays and TeV gamma rays from several shell-type SNRs. 
\cite{koyama95} discovered synchrotron X-rays from the 
shell of SN~1006, which indicates electrons are accelerated 
up to multi-TeV energies. 
This finding was followed by detections of 
synchrotron X-rays from other SNRs, including RX~J1713.7$-$3946 
(e.g. \citealt{koyama97,slane01}). 
Further evidence for multi-TeV particles 
(electrons and/or protons) has been provided by discovery of 
TeV gamma rays from some SNRs, such as Cassiopeia~A \citep{aha01} 
or RX~J1713.7$-$3946 \citep{muraishi00}, although their spectral parameters 
and morphologies were not well determined due to the limited sensitivity of TeV observatories. 
Subsequently, high quality morphological 
 and spectral studies have been performed by H.E.S.S.   
(e.g. \citealt{aha04,aha06,aha07}). 
These pioneering measurements by the H.E.S.S. telescope, 
together with the high resolution X-ray data,  
have enabled direct comparison 
of X-ray and TeV gamma-ray data. 

The shell-type SNR RX~J1713.7$-$3946 (also known as G347.3$-$0.5), 
is one of the best-studied SNRs from which both non-thermal X-rays 
and TeV gamma rays are detected. This SNR was discovered in 
soft X-rays during the {\it ROSAT} All-Sky Survey \citep{pfe96}. 
The {\it ASCA} satellite, with wider energy coverage than that of {\it ROSAT}, 
revealed that the X-ray spectrum is featureless and can be best 
interpreted as synchrotron emission from very high energy electrons 
in the TeV regime \citep{koyama97,slane99}. 
The X-ray spectrum was well fitted with a power-law function 
of photon index $\Gamma = 2.2$--$2.4$ and  interstellar absorption 
column density $N_{\rm H} = 0.6$--$0.8 \times 10^{22}~{\rm cm}^{-2}$ without any 
observable evidence for a thermal emission component. 
Subsequent observations by {\it Chandra} and {\it XMM-Newton} 
have unveiled structure with a complex network of bright filaments and knots, 
in the western part of the SNR \citep{uchi03,laz04,cassam04,hiraga05}. 

TeV gamma-ray emission from RX~J1713.7$-$3946 was first reported by 
the CANGAROO collaboration in 1998 \citep{muraishi00}, and 
confirmed by the subsequent observations with CANGAROO-II in 2000 
and 2001 \citep{enomoto02}. 
Later, the H.E.S.S. collaboration obtained  a resolved  image of the source 
in TeV gamma rays \citep{aha04} showing that 
the gamma-ray emission from RX~J1713.7$-$3946 arises mainly in the shell. 
These observations revealed  
a striking correlation between the X-ray and the gamma-ray images, 
which indicates a strong connection between the physical processes responsible for X-ray
and TeV gamma-ray emission components \citep{aha06}.  Based on the spectral and morphological 
information, they discussed two possible gamma-ray emission scenarios, 
one where gamma rays are generated by inverse Compton scattering of 
accelerated electrons with diffuse radiation  fields  (the so-called leptonic scenario) 
and the other where the decay of  secondary $\pi^{0}$-mesons is  responsible for gamma rays 
(hadronic scenario). 
The later observations with H.E.S.S.  revealed that the flux  extends to  
30~TeV and, likely, beyond,  which implies particle acceleration up to 
energies  well above 100~TeV for either 
model \citep{aha07} . 

Most recently, our X-ray observations using {\it Chandra} and {\it Suzaku} have 
provided important clues for understanding the acceleration process
in the SNR. 
From a series of observations 
of the northwest part of the SNR 
with {\it Chandra} in 2000, 2005 and 2006, we 
discovered that compact regions of the northwest (NW) shell are variable in flux 
on a one-year time scale \citep{uchi07}. The fast variability was 
interpreted as one-year scale acceleration and synchrotron cooling of 
electrons with amplified magnetic fields of order of 1~mG. 
Such a large magnetic field in compact regions
strongly favors $\pi^0$-decay emission as the origin of TeV gamma rays. 
Also, thanks to the wide-band coverage of {\it Suzaku} and its low background level, 
 we were able to measure a hard X-ray spectrum up to 40~keV from 
the southwest portion of RX~J1713.7$-$3946 with a clear indication of 
a high-energy cutoff in the synchrotron spectrum (\citealt{takahashi07}, hereafter Paper I). 
Combined with the upper limit on a shock speed of 
$4500~{\rm km}~{\rm s}^{-1}$ placed by {\it Chandra}, 
the cutoff energy determined by the  {\it Suzaku} observation
of the southwest part  indicates that particle 
acceleration within the SNR shock is so efficient that it approaches the theoretical 
limit corresponding to  the so-called Bohm diffusion regime (e.g. \citealt{malkov01}).

In this paper, we present results of mapping observations 
of RX~J1713.7$-$3946 with {\it Suzaku}, 
which covers about two-thirds of the SNR region with 11 pointings. 
The low background level of the Hard X-ray Detector (HXD) enables us to detect hard X-ray emission 
up to $\sim 40~{\rm keV}$ from each of the pointings. 
At the same time, its small field-of-view (FoV) of $\sim 25^{\prime} \times 25^{\prime}$ FWHM gives us information about the spatial distribution of hard X-ray emission and 
spectral differences from region to region. 
Thanks to its low instrumental background and large effective area,  
the other detector system aboard {\it Suzaku}, the X-ray Imaging Spectrometer (XIS), 
also uncovers new observational 
facts, such as spectral features below 10~keV and the morphology of relatively dim 
regions left unclear in previous studies by {\it ASCA}, {\it Chandra}, and 
{\it XMM-Newton}. By combining the XIS and HXD spectra summed over the data from 
all the pointings, we show a wide-band X-ray spectrum (0.4--40~keV) with quite high 
statistics, with which we investigate not only the existence of a cutoff, but also 
its shape. We then compare the cutoff shape obtained with theoretical predictions. 

In Section 2, we describe our {\it Suzaku} observations and the data reduction 
procedures. Analysis and results of HXD and XIS data are shown in \S3.1 and 
\S3.2, respectively. We present the wide-band spectrum by connecting the XIS 
and the HXD data in \S3.3. A detailed study regarding the cutoff structure is 
also given there. Section 4 is devoted to multi-wavelength 
spectral and morphological studies. 
The results obtained are discussed in 
the following section, and the results are 
finally summarized. 
Throughout this paper, we assume that the  distance to RX~J1713.7$-$3946 
is close  to  1~kpc as proposed by \cite{koyama97} based on the $N_{\rm H}$ value. 
A similar distance has been claimed based on the NANTEN CO data \citep{fukui03,moriguchi05}. 
The typical age of the remnant for such a distance is estimated of order of 1000~yr, which
can be an indication of  association of RX~J1713.7$-$3946  with an explosion 
in A.D. 393 as proposed by \cite{wang97}. 

\section{SUZAKU OBSERVATIONS AND DATA REDUCTION}
The {\it Suzaku} observatory \citep{mitsuda07} is the fifth Japanese X-ray astronomy satellite,  
jointly developed by Japan and the US. 
Its scientific payload consists of two co-aligned detector systems, 
the XIS \citep{koyama_xis07} and the HXD \citep{takahashi_hxd07,kokubun07}. 
The XIS consists of four X-ray CCD cameras which are located in the foci of 
X-ray telescopes (XRT; \citealt{serlem07}). Three of the XIS sensors are front-illuminated 
(FI; 0.4--12~keV) CCDs and the other is back-illuminated (BI 0.2--12~keV). 
The non-imaging, collimated hard X-ray instrument, the HXD, covers the
10--600~keV bandpass. Two main detector units, silicon PIN diodes and 
GSO scintillators, are buried at the bottom of  well-type 
active shield of BGO. The former covers the lower energy band of 10--60~keV, while 
the latter detects higher energy photons of 40--600~keV. 

\begin{figure}
\epsscale{1.0}
\plotone{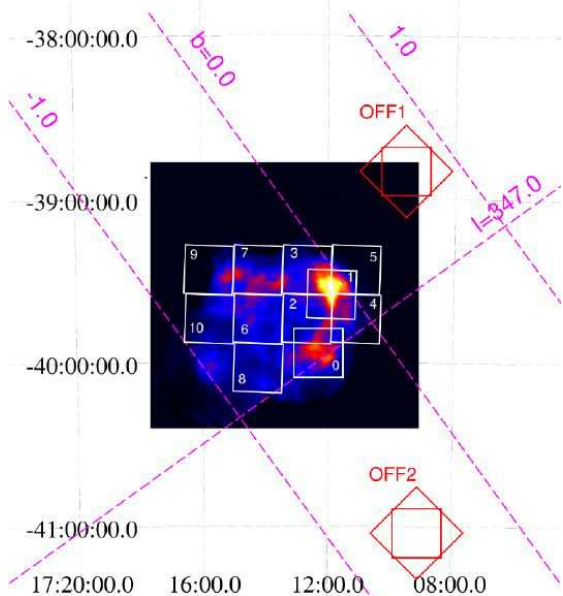}
\caption{{\it Suzaku} FoV of each observation of the RX~J1713.7$-$1713 region overlaid on the 
{\it ASCA} GIS image (1--5~keV) taken from \cite{uchi05}. The small squares corresponds to the FoV of the XIS. 
the outer diamonds drawn for each ``OFF'' pointing is the rough shape of the FOV (50\% of the effective area) 
of the HXD PIN. The numbers indicated in the XIS FoV are pointing IDs used throughout this paper.}
\label{fig:obs_fov}
\end{figure}

We performed 11 pointing observations of RX~J1713.7$-$3946 with {\it Suzaku}. 
The observation log is summarized in Table~1, 
and the pointing position of each observation is shown in Figure~\ref{fig:obs_fov}.   
The southwest part of the SNR, labeled as Pointing 0, was observed in 2005 during 
the Performance Verification phase, while the other ten observations 
were performed in 2006 during the {\em Suzaku} AO1 phase. Since the SNR is 
located on the Galacic plane, it is of importance to check the hard X-ray background 
associated with the Milky Way. We therefore observed two nearby background regions 
containing no bright X-ray point sources. 
The pointing positions of these ``OFF'' observations are shown with red squares 
in Figure~\ref{fig:obs_fov}. 
The XIS was operated in the normal full-frame clocking mode without spaced-row charge 
injection during all the observations. 
Since the results from Pointing 0, together with those from the OFF observations, 
are already reported in Paper I, we do not give a detailed description 
on the analysis and results for these data.  

We used data products from the pipeline processing version 1.2. 
For the XIS analysis, we retrieved ``cleaned event files'' which are screened 
using standard event selection criteria. 
For the 2006 data, we recalculated the values of pulse invariant (PI) 
and the grade values since incorrect CALDB is applied to the pipeline 
processing of these data as 
announced by the {\it Suzaku} instrument teams. 
We further screened the cleaned events 
with following criteria as recommended by the {\it Suzaku} instrument teams 
-- (1) cut-off rigidity larger than 6~GV and (2) elevation angle from the 
Earth rim larger than $10^{\circ}$. For the HXD data, ``uncleaned event files'' 
were screened using standard event screening criteria. The exposure times 
after these screenings are shown in Table~1. 
Due to unstable operation of 16 PIN diodes installed in
Well-counter units, W00--W03; hereafter W0, 
the bias voltage for these diodes was reduced to 400~V from the nominal 
voltage of 500~V on 26 May 2006.   
On 2006 October 4, the bias voltage of 16 more PIN diodes 
(the PIN diodes in the Well-counter units of W10--W13; W1) 
was reduced to 400~V for the same reason. 
The reduction of the PIN diode bias voltage leads to a decrease of their effective area and also affects their energy response. 
Since the current response matrices do not include these effects, 
only the PIN diodes with a bias voltage of 500~V are utilized in the following analysis. 
Throughout this paper, the data reduction and analysis are performed using HEADAS 6.2 
and the spectral fitting is done with XSPEC 11.3.2. 

\begin{deluxetable*}{ccccc}
\tablecolumns{5}
\tablewidth{0pc}
\label{tab:obs}
\tablecaption{Summary of the {\it Suzaku} observations of RX~J1713.7$-$3946}
\tablehead{
Pointing ID & Obs. ID & Coord. (J2000) & Exposure & Date \\ 
&  & ($\alpha_{\rm J2000}$,  $\delta_{\rm J2000}$) & XIS/HXD & \\
& &  &  [ks] & 
}
\startdata
0 & 100026010 & ($17^{\rm h}12^{\rm m}17^{\rm s}.0$, $-39^{\rm d}56^{\rm m}11^{\rm s}$) & 55/48 & 26/9/2005\\
1 & 501063010 & ($17^{\rm h} 11^{\rm m} 51^{\rm s}.5$, $-39^{\rm d} 31^{\rm m} 13^{\rm s}$) & 17/17 & 11/9/2006\\
2 & 501064010 & ($17^{\rm h} 12^{\rm m} 38^{\rm s}.0$, $-39^{\rm d} 40^{\rm m} 14^{\rm s}$) & 18/22 & 11/9/2006\\
3 & 501065010 & ($17^{\rm h} 12^{\rm m} 38^{\rm s}.2$, $-39^{\rm d} 22^{\rm m} 15^{\rm s}$) & 19/18 & 11/9/2006\\
4 & 501066010 & ($17^{\rm h} 11^{\rm m} 04^{\rm s}.5$, $-39^{\rm d} 40^{\rm m} 10^{\rm s}$) & 19/21 & 12/9/2006\\
5 & 501067010 & ($17^{\rm h}11^{\rm m} 05^{\rm s}.1$, $-39^{\rm d} 22^{\rm m} 10^{\rm s}$) & 16/19 & 12/9/2006\\
6 & 501068010 & ($17^{\rm h} 14^{\rm m} 11^{\rm s}.6$, $-39^{\rm d} 40^{\rm m} 14^{\rm s}$) & 20/19 & 13/9/2006\\
7 & 501069010 & ($17^{\rm h} 14^{\rm m} 11^{\rm s}.4$, $-39^{\rm d} 22^{\rm m} 15^{\rm s}$) & 12/11 & 19/9/2006\\
8 & 501070010 & ($17^{\rm h} 14^{\rm m} 11^{\rm s}.8$, $-39^{\rm d} 58^{\rm m} 14^{\rm s}$) & 19/20 & 19/9/2006\\
9 & 501071010 & ($17^{\rm h} 12^{\rm m} 17^{\rm s}.6$, $-39^{\rm d} 18^{\rm m} 50^{\rm s}$) & 16/15 & 20/9/2006\\
10 & 501072010 & ($17^{\rm h} 15^{\rm m} 44^{\rm s}.5$, $-39^{\rm d} 40^{\rm m} 10^{\rm s}$) & 15/15 & 5/10/2006\\
\hline
OFF1 & 100026020 & ($17^{\rm h} 09^{\rm m} 31^{\rm s}.9$, $-38^{\rm d} 49^{\rm m} 24^{\rm s}$) & 28/24 & 25/9/2005\\
OFF2 & 100026030 & ($17^{\rm h} 09^{\rm m} 05^{\rm s}.1$, $-41^{\rm d} 02^{\rm m} 07^{\rm s}$) & 30/28 & 28/9/2005
\enddata
\end{deluxetable*}

\section{ANALYSIS AND RESULTS}
\subsection{HXD Data Analysis}\label{subsec:hxd_ana}

\subsubsection{Spectral Analysis}\label{subsec:rxj1713_hxd_spec}
The HXD PIN spectrum from each pointing was constructed and compared 
with the background model estimated for the each observation period. 
In the analysis below, the non-Xray background (NXB) model 
\citep{watanabe07} provided 
by the HXD team is used for the background generation. 
Since the NXB model does not include the contributions from 
the cosmic X-ray background (CXB), 
a simulated CXB spectrum was added to the NXB model. 
Specifically, based on the reanalysis of the data from {\it HEAO-1} 
observations in the 1970's \citep{gruber99}, the CXB spectrum was modeled as 
\begin{eqnarray}
\frac{dN}{d\varepsilon} = 7.9~{\varepsilon_{\rm keV}}^{-1.29} \exp \left( -\frac{\varepsilon_{\rm keV}}{\varepsilon_{\rm p}} \right)~{\rm ph}~{\rm s}^{-1}~{\rm keV}^{-1}~{\rm cm}^{-2}~{\rm str}^{-1}, 
\end{eqnarray}
where $\varepsilon_{\rm keV} = \varepsilon/1~{\rm keV}$ and $\varepsilon_{\rm p} = 41.13$. 
We estimated the CXB signal in each HXD PIN spectrum using the response 
matrix for spatially uniform emission, {\tt ae\_hxd\_pinflat\_20060809.rsp}, 
{\tt ae\_hxd\_pinflat123\_20060809.rsp}, and {\tt ae\_hxd\_pinflat23\_20060809.rsp}. 
The contribution from the CXB flux to the detected count 
rate is estimated to be  $\sim 5$\% of the NXB. 

\begin{figure}
\epsscale{1.0}
\plotone{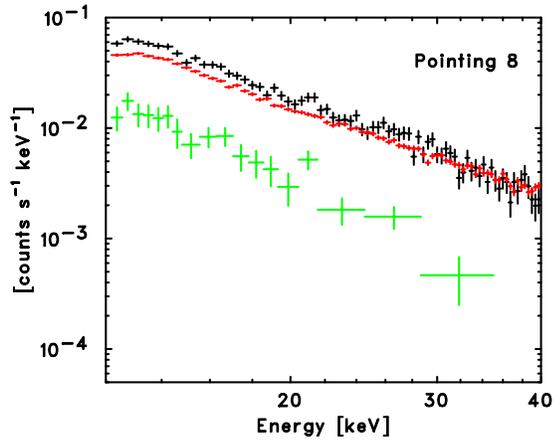}
\caption{The HXD PIN spectrum obtained from the observation of Pointing 8. 
The data points in black show the raw spectrum, red points represent 
the background model (NXB+CXB), and green points are the background-subtracted spectrum. }
\label{fig:hxd_spec_p8}
\end{figure}

Figure~\ref{fig:hxd_spec_p8} shows the HXD PIN spectrum obtained 
from Pointing 8, where a clear detection of hard X-rays can be seen. 
Likewise, the HXD PIN detected signals from all other pointings 
with a count rate of 20--50\% of the NXB. 
We fitted the background-subtracted spectra with a simple power 
law model: $dN/d\varepsilon \propto \varepsilon^{-\Gamma}$. 
Since the extended nature of the source does not cause any spectral 
steepening or flattening (Paper I), we used the point-source 
response matrix at the XIS-nominal position, 
{\tt ae\_hxd\_pinxinom\_20060814.rsp}, {\tt ae\_hxd\_pinxinom123\_20060814.rsp}, 
and {\tt ae\_hxd\_pinxinom23\_20060814.rsp}. 
Table~2 gives the best-fit parameters with the 
statistical errors at 90\% confidence level. The obtained photon indices 
are generally larger than those obtained from the corresponding XIS spectra (see below).  
This difference indicates that a spectral cutoff is not unique to the SW region (Paper I), but a common feature 
throughout the remnant. 

In order to confirm the results obtained above, we evaluated the systematic 
errors due to uncertainties in the NXB modeling. 
As described in \cite{mizuno06}, the current reproducibility of the 
NXB model is $\sim 5$\%. 
Therefore, we examined how much 
the values of the photon index change 
by increasing or decreasing the background model by 5\%. 
The systematic errors were found to be 
smaller than the statistical errors indicated in 
Table~2. 

\begin{figure}
\epsscale{1.0}
\plotone{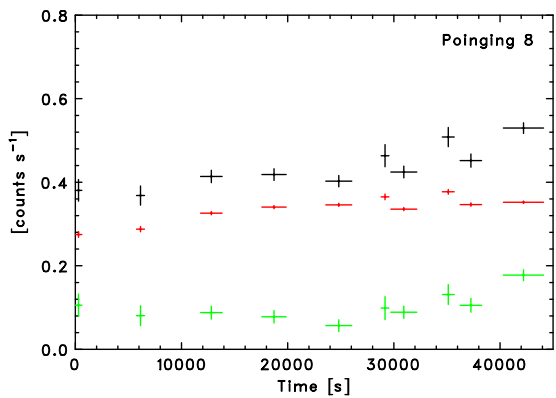}
\caption{The HXD PIN light curve during the observation of Pointing 8. 
Events with energy from 12~keV to 40~keV are selected. 
The black, red, and green points show the raw, background (NXB), 
and background-subtracted data, respectively.}
\label{fig:hxd_lc_p8}
\end{figure}

The systematic errors due to the misestimation of NXB were 
examined in another way. 
Considering the physical size of 
this target, the emission from the remnant  
should be constant during the observations. 
Therefore, background-subtracted lightcurves should 
be constant during a observation. 
Although the lightcurves shown 
are almost constant within statistical errors, 
the background-subtracted count rate becomes
higher when the total count rate increases for 
Pointings 0 and 8. 
The light curves for Pointings 0 and 8 are shown 
in Paper I and Figure~\ref{fig:hxd_lc_p8}, respectively. 
Since this behavior is thought to be caused by 
misestimation of the NXB, we examined how much the 
photon indices change with and without those time regions. 
When we discard the time region which corresponds to 
the last bin of Figure~\ref{fig:hxd_lc_p8}, the photon index 
changes by $\Delta \Gamma \simeq 0.2$ from the values in 
Table~2. 

Diffuse emission from the Galactic plane also can affect 
the spectra. However, no emission above the 5\% level of the NXB 
was detected from the OFF pointings (Paper I). 
Moreover, when the 
excess counts marginally detected from the OFF pointings 
are added to the background spectrum for each pointing, 
the fitting result agrees within the statistical errors.

\begin{deluxetable*}{ccccc}
\label{tab:hxd_fit}
\tablecolumns{5}
\tablewidth{0pc}
\tablecaption{Power-law fitting to the HXD PIN spectra\tablenotemark{a}}
\tablehead{
Pointing ID & $\Gamma$ & $F_{10-40~{\rm keV}}$ & ${\chi_\nu}^2~(\nu)$ & Used Units \\ 
\colhead{} & \colhead{} &   (mCrab)  & \colhead{}  & \colhead{} 
 }
\startdata
0 & $3.2\pm0.2$ & $2.5\pm0.1$ & 1.15 (36) & W0--3 \\
1 & $3.3\pm0.2$ & $3.2\pm0.2$ & 0.91 (18) & W1--3 \\
2 & $3.0\pm0.3$ & $3.2\pm0.2$ & 0.84 (21) & W1--3 \\
3 & $3.4\pm0.5$ & $2.0\pm0.2$ & 0.58 (17) & W1--3 \\
4 & $2.9\pm0.3$ & $3.5\pm0.2$ & 0.77 (23) & W1--3 \\
5 & $3.2\pm0.5$ & $1.7\pm0.2$ & 0.76 (17) & W1--3 \\
6 & $2.9\pm0.3$ & $2.9\pm0.2$ & 0.72 (20) & W1--3 \\
7 & $3.9^{+1.4}_{-1.1}$ & $1.2^{+0.3}_{-0.2}$ & 0.42 (9) & W1--3 \\
8 & $2.6^{+0.5}_{-0.4}$ & $2.2\pm0.2$ & 0.87 (19) & W1--3 \\
9 & $3.0^{+1.2}_{-1.0}$ & $1.1\pm0.2$ & 0.86 (12) & W1--3 \\
10 & $4.4^{+1.6}_{-1.2}$ & $1.0\pm0.2$ & 0.20 (8) & W2, 3
\enddata
\tablenotetext{a}{Errors represent 90\% confidence.}
\end{deluxetable*}

\subsubsection{Spatial Distribution of Hard X-ray Emission}\label{subsec:hxd_spatial}
Using the spectral parameters obtained for the 11, pointings we attempt to
reconstruct the spatial distribution of the hard X-ray 
emission.
Since we expect spatial variation of both brightness and spectral index, 
we use a Monte Carlo simulator, {\tt simHXD} 
\citep{terada05}, for the estimation. 
We input the spatial distribution of brightness and spectral index 
into {\tt simHXD}, and compare the simulated spectra with 
observed ones shown in \S\ref{subsec:rxj1713_hxd_spec}. 

\begin{figure}
\epsscale{1.0}
\plotone{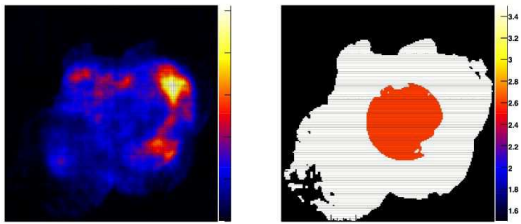}
\caption{The emission distribution (left) and the photon index 
distribution (right) adopted in the simulations. 
The emission distribution is the {\it ASCA} image in the energy 
range of 1--5~keV.}
\label{fig:sim_dist4}
\end{figure}

\begin{figure}
\epsscale{1.0}
\plotone{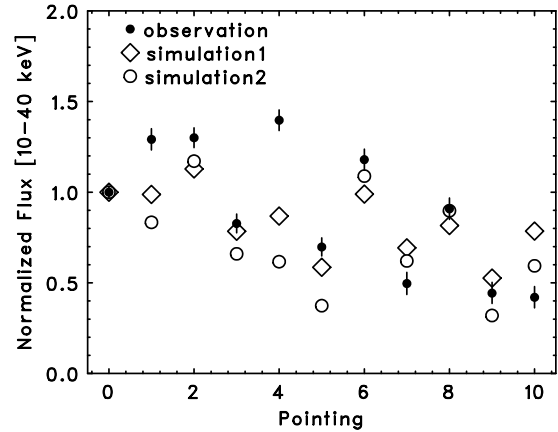}
\caption{Comparison of detected flux obtained from the observations, 
{\it simulation1}, and {\it simulation2}. Each data point is normalized to the value of 
Pointing 0. The error bars indicate $1\sigma$ statistical errors.}
\label{fig:norm_hikaku}
\end{figure}

Firstly we simulated by taking a simple model in which 
the brightness distribution is same as the {\it ASCA} image shown 
in Figure~\ref{fig:sim_dist4} and the spectral index is 
constant at $\Gamma = 3.0$ throughout the SNR ({\it simulation1}). 
By using the {\it ASCA} image as an input to {\tt simHXD}, 
we simulated hard X-ray spectra in 10--40 keV, which are expected to be
observed by the PIN in each pointing. 
Figure~\ref{fig:norm_hikaku} compares the flux
from the observations and the simulations. 
In this figure, each value is normalized to that of Pointing 0. 
Here, the systematic error for each observational data point is 
$\sim 20$\% if we use the 5\% of the background 
as systematic errors. The observations and the simulations 
appear consistent with each other. Therefore, we expect 
the brightness distribution above 10~keV is not drastically different 
from the distribution below 10~keV. 

The largest discrepancy between 
the observation and the simulation is found for 
Pointing 4. 
In this pointing, there is a known point source 
at the corner of the HXD-PIN field of view. 
The source is listed in the Second IBIS/ISGRI Soft Gamma-Ray 
Survey Catalog \citep{bird06} as IGR~J17088$-$4008. 
According to \cite{bird06}, 
the average flux of this source is  
$1.1\pm0.2$~mCrab in 20--40~keV and $2.2\pm0.3$~mCrab 
in 40--100~keV. 
It is noted that the source is not bright enough for IBIS/ISGRI 
to determine spectral parameters. 
The transmission of fine collimators of the HXD for the source 
is estimated to be $\sim 0.05$. The estimated 
count rate of this source corresponds to $\sim 2$\% of 
the detected signals from Pointing 4. 
However, the variability 
of this source could increase the count rate to a non-negligible 
level. The observation and the simulation become consistent 
if the point source was 10 times brighter than its 
average value during the {\it Suzaku} observation.  
Also, the angular response of the HXD near the edge of the 
FoV can have some uncertainties since calibrations for a source 
with such a large offset angle is difficult. 
The uncertainties can affect our estimate of the 
contamination from IGR~J17088$-$4008.  
At this moment, we cannot conclude that the large difference between 
the observation and the simulation is due to the spatial 
distribution of the SNR emission or 
the point source in the FoV.

Next we tried another simulation taking into account 
the spatial distribution of spectral indices ({\it simulation2}). 
We adopt a ``toy model'' 
shown in Figure~\ref{fig:sim_dist4} since the results 
of the spectral analysis presented in Table~2 
suggest that the hard X-ray spectrum may be flatter in the inner region 
of the SNR than near the rim. 
In the model, the photon index of the inner region is set to $\Gamma = 2.6$ 
and that of the rim region is set to $\Gamma = 3.5$. 
The {\it ASCA} image was used to provide the brightness distribution 
at 10~keV. 

\begin{figure}
\epsscale{1.0}
\plotone{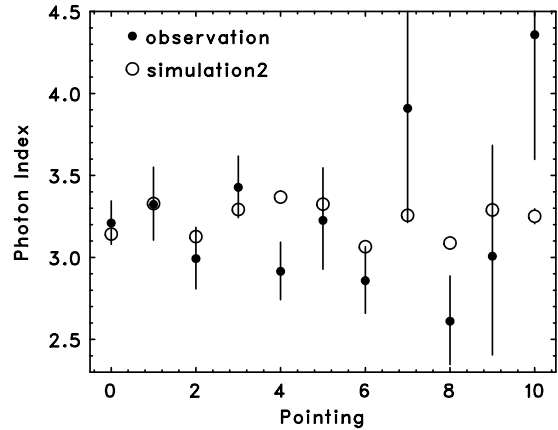}
\caption{Comparison of photon indices obtained from the observations 
and {\it simulation2}. 
The error bars 
correspond to $1\sigma$ statistical errors.}
\label{fig:index_hikaku}
\end{figure}

Comparisons of the data and the simulation results  
are shown in Figure~\ref{fig:norm_hikaku} and 
Figure~\ref{fig:index_hikaku} for the flux and photon index, 
respectively. 
The detected flux 
obtained from {\it simulation2} is somewhat similar to 
that from {\it simulation1}, and the simulation data generally 
follow the observational data. 
As for the distribution of photon indices, the toy model gives 
a similar distribution 
to the observational results for the western portion of the SNR. 
Obtaining a better fit for the eastern region (Pointing 7, 8, 
and 10) may require more complex assumptions than the toy model 
for {\it simulation2}. 


\subsection{XIS Data Analysis}\label{subsec:xis_ana}
\subsubsection{Image Analysis}
\begin{figure}
\epsscale{1.0}
\plotone{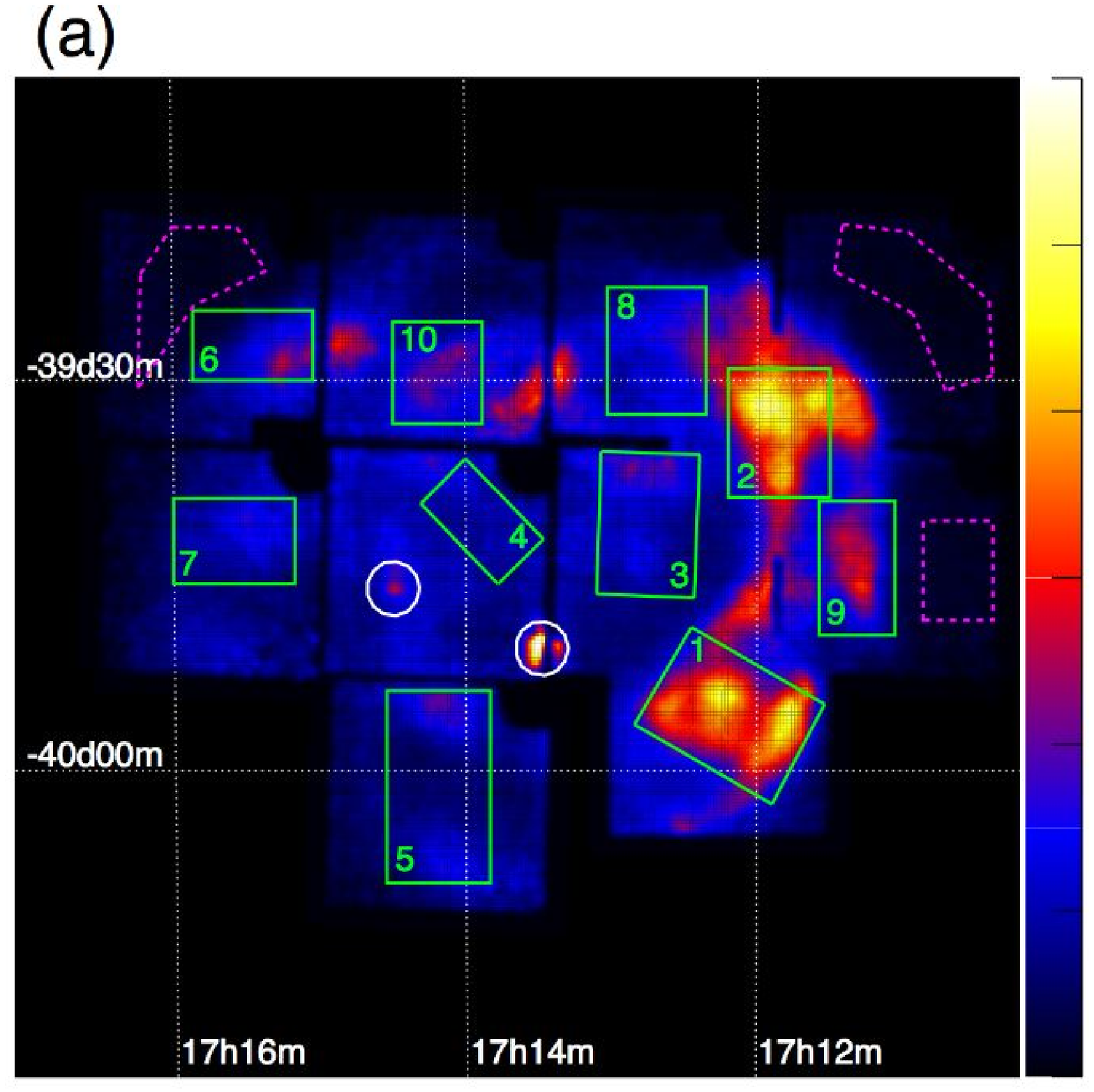}
\plotone{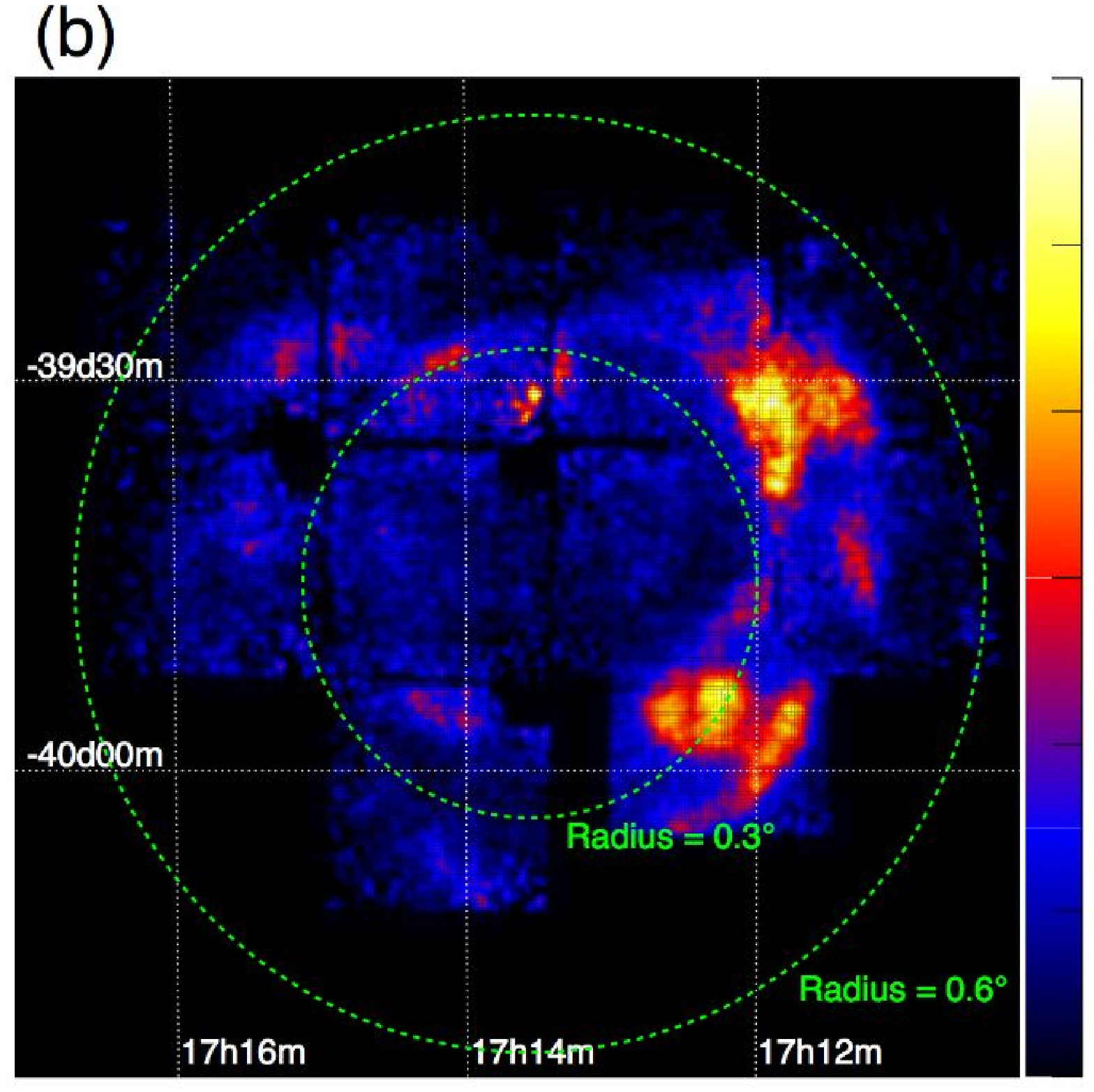}
\caption{The XIS (XIS 0+2+3) mosaic images of RX~J1713.7$-$3946 in the energy band 1--5~keV (a) and 
5--10~keV (b). North is up and east is to the left. 
The color scale indicates count rate on a linear scale. The green regions in the top panel 
are those used for spectral analysis. The dashed magenta polygons correspond to the regions used to 
extract the background spectrum. The positions of the two point sources, 1WGA~J1714.4$-$3945 
and 1WGA~J1713.4$-$3949 are shown with white circles. The dashed green circles represent 
radii of $0.3^\circ$ and $0.6^\circ$ around the center of the SNR 
($\alpha_{\rm J2000} = 17^{\rm h}13^{\rm m}33.6^{\rm s}$,  
$\delta_{\rm J2000} = -39^{\rm d}45^{\rm m} 36^{\rm s}$), which correspond to 
the vertical dashed lines in Figure~\ref{fig:xis_radial}.}
\label{fig:xis_mosaic_image}
\end{figure}

Figure~\ref{fig:xis_mosaic_image} shows mosaic images of RX~J1713.7$-$3946, 
constructed using the data from XIS0, 2, and 3 (FI-CCDs). 
The upper panel shows the soft band image in 1--5~keV and the lower panel 
shows the hard band image in 5--10~keV. 
Both images are smoothed with a Gaussian of $\sigma = 0^{\prime}.3$.  
Instrumental background signals 
are subtracted from both images. 
The signal to background ratio in the hard band is smaller than that in the soft band by 
one order of magnitude. Thus, the background must be carefully subtracted. 
We utilized the Night Earth Background Database consisting of event data 
obtained when the satellite is looking at the night earth and the 
non X-ray background becomes dominant.
After subtracting the background, the vignetting effects of the XRTs were 
corrected by means of the simulation program called {\tt xissim} \citep{ishisaki07}. 
In the program, an image from a flat field can be simulated by a Monte Carlo method. 

\begin{figure}
\epsscale{1.0}
\plotone{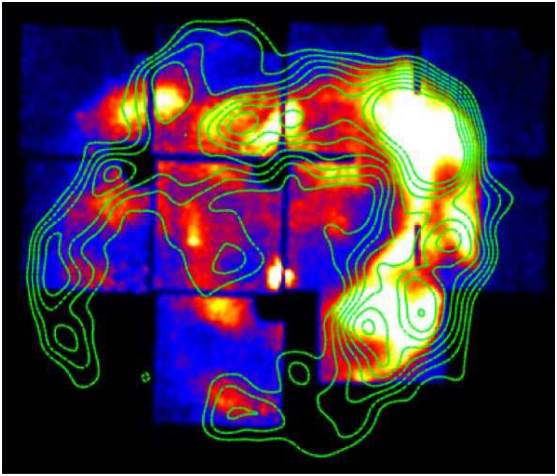}
\caption{Comparison of the {\it Suzaku} XIS image and the gamma-ray image by the H.E.S.S. telescope (contours) taken 
from \cite{aha07} shown with color scale and green contours, respectively. The XIS image is same as 
Figure~\ref{fig:xis_mosaic_image} (a) but the scale is changed to stress the similarity of the two images.}
\label{fig:xis_hess_image}
\end{figure}

As is seen in Figure~\ref{fig:xis_mosaic_image}, the {\it Suzaku} XIS has covered most  of the remnant. 
Thanks to little stray-light contamination of the XRTs and low background 
level of the XIS, high quality images are obtained even in the high energy band above 5~keV.  
The double shell structure revealed by {\it XMM-Newton} is clearly seen in the XIS 
images. 
In addition to the bright structures of the western part, 
the XIS revealed detailed morphology 
of the dim parts of the remnant. 
The dim structures are highlighted in 
Figure~\ref{fig:xis_hess_image}, which is the same as 
Figure~\ref{fig:xis_mosaic_image} (a) but displayed with a different color scale. 
In this figure, the contours of the H.E.S.S. gamma-ray image (from \citealt{aha07}) 
are overlaid for comparison. As is clearly seen, not only the bright rims but also 
the eastern portion shows a striking similarity between the two 
energy regimes. This correlation is discussed more quantitatively in \S\ref{subsec:mws}.

Two point sources seen in the soft band image 
are listed in the {\it ROSAT} bright source catalogue. One is 
1WGA~J1714.4$-$3945, which was associated with a Wolf-Rayet star by \cite{pfe96}. 
The other is 1WGA~J1713.4$-$3949, which is located between the two FoVs of 
Pointing 2 and 6. This source has been suggested to be the neutron star associated 
with the SNR by \cite{laz03}.

\begin{figure}
\epsscale{1.0}
\plotone{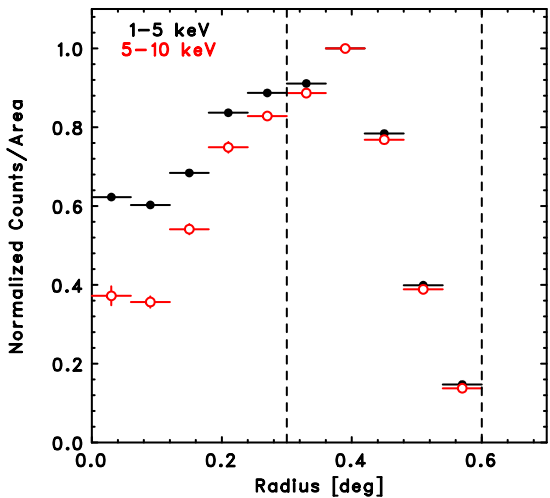}
\caption{Radial profiles around the center of the SNR 
($\alpha_{\rm J2000} = 17^{\rm h}13^{\rm m}33.6^{\rm s}$,  
$\delta_{\rm J2000} = -39^{\rm d}45^{\rm m} 36^{\rm s}$) 
in the two energy bands. Each profile is normalized to its peak value for 
direct comparison.The radii corresponding the vertical dashed lines are shown in 
Figure~\ref{fig:xis_mosaic_image} (b) with the green dashed circles.}
\label{fig:xis_radial}
\end{figure}

Comparing the soft-band and hard-band images provides information about 
the spatial variations of the spectral properties. 
At first sight, the bright structures are very similar in the soft band  and hard band images. 
In order to compare the images in more detail, 
a radial profile around the center of SNR
($\alpha_{\rm J2000} = 17^{\rm h}13^{\rm m}33^{\rm s}.6$,  
$\delta_{\rm J2000} = -39^{\rm d}45^{\rm m} 36^{\rm s}$) is presented 
in Figure~\ref{fig:xis_radial}, 
where circular regions of $2^{\prime}.1$ radius centered on the 
two point sources are excluded. 
The hard band to soft band ratio is significantly different 
between the bright outer region part and the interior. This difference suggests a 
corresponding difference in spectral properties. 

\subsubsection{Spectral Analysis}\label{subsec:xis_spec}
As already reported in Paper I, the XIS data reveal
a cutoff below 10~keV in the spectrum of the SW rim. 
The XIS spectra from the various regions should reveal whether 
the cutoff varies across the remnant. 
We extracted spectra from the regions shown in 
Figure~\ref{fig:xis_mosaic_image} with the green squares. 
The region labeled as ``1' (Region 1) is the same one that 
we used for the spectral analysis in Paper I. 
The background spectrum needs to be accumulated from nearby 
regions since RX~J1713.7$-$3946 is located almost on 
the Galactic plane. The OFF observation data cannot 
be used for background estimation other than for Region 1. 
This is because the contamination on 
the optical blocking filters of the XIS \citep{koyama_xis07} significantly 
changed the detector response to low-energy X-rays between 
2005 and 2006. 
For Region 1--10 we therefore accumulated a background spectrum 
from the regions indicated with the magenta polygons in 
Figure~\ref{fig:xis_mosaic_image}. 
In the spectral fitting discussed below, the standard RMF files version 2006-02-13 
were used, whereas ARF files were produced using {\tt xissimarfgen}. 
Each spectrum is binned so that each bin contains at least 300 counts. 
After the binning, we ignore those bins whose energy is smaller than 
0.4~keV or larger than 12~keV. We also excluded bins between 
1.7 and 1.9~keV because there exist large systematic uncertainties in 
the response matrices. 
Finally, we co-added the spectra, RMF files, and ARF files from the FI 
chips to produce a single data set for the XIS.

\begin{figure}
\epsscale{1.0}
\plotone{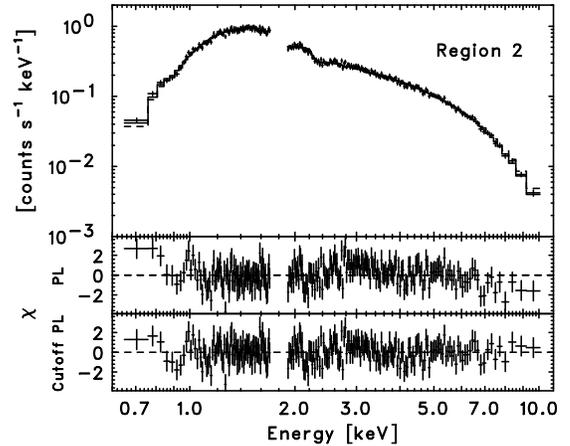}
\caption{The XIS (XIS 0+2+3) spectrum from Region 2. The lower panels show the residuals 
when the spectrum is fitted with a power law and a power law with an exponential cutoff.}
\label{fig:xis_spec_reg2}
\end{figure}

First, we fitted all the spectra with a simple power law absorbed 
by the interstellar medium. The results are summarized in 
Table~3. This model yields acceptable 
fits for most of the spectra.
However, the fits to the spectra of Region 1, 2, and 6 
are not acceptable even at the 99\% confidence level. 
We show the spectrum of Region 2 with the best-fit power-law and the residuals in Figure~\ref{fig:xis_spec_reg2}. 
Although not so drastic as Region 1 which is shown 
in Paper I, a correlated pattern to the residuals 
can be seen for the Region 2. A similar pattern is also seen 
in the data of Region 6. 

We the fitted all the spectra with a cutoff power law. This function 
gives acceptable fits for all regions, and consistently gives better values 
of $\chi_{\nu}^2$. In Figure~\ref{fig:xis_spec_reg2}, we plot the 
residuals for the spectrum of Region 2. 
These results, together with the steeper spectra detected 
with HXD, indicate the existence of a cutoff 
somewhere between the bandpasses of the XIS and HXD. 

Although statistically rejected for some regions, the results obtained 
with the power-law fits are consistent with previous studies 
with {\it ASCA}, {\it Chandra}, and {\it XMM-Newton} 
\citep{koyama97, slane99, uchi03, cassam04, hiraga05}.  
The results with {\it Suzaku} suggest that the photon index is larger  
and the absorption is smaller for the inner regions than for the outer regions. 
This tendency agrees with the radial profile shown in Figure~\ref{fig:xis_radial}. 
The difference of $N_{\rm H}$ between the western bright spots 
and the inner region is 
$\Delta N_{\rm H} \simeq 0.3$--$0.4 \times 10^{22}~{\rm cm}^{-2}$,  
which is consistent with the results from {\it XMM-Newton}. 
According to the discussion by \cite{hiraga05}, there could be a correlation between 
the difference of $N_{\rm H}$ and the presence of the molecular clouds 
in the western part of the SNR detected with the NANTEN telescope 
\citep{fukui03,moriguchi05}. 

\begin{deluxetable*}{cccccccccccc}
\label{tab:xis_fit}
\tabletypesize{\footnotesize}
\tablecaption{Model fitting to the XIS spectra\tablenotemark{a}}
\tablewidth{0pt}
\tablehead{
\colhead{Region} & \colhead{Area} & \multicolumn{4}{c}{Power law} & \colhead{} & \multicolumn{5}{c}{Cutoff power law} \\
\cline{3-6} \cline{8-12} 
\colhead{} & \colhead{} & $N_{\mathrm H}$ & $\Gamma$ & $F_{1-10~{\rm keV}}$\tablenotemark{b} & ${\chi_\nu}^2 (\nu)$ & \colhead{} & $N_{\rm H}$ & $\Gamma$ & $\varepsilon_c$ &$F_{1-10~{\rm keV}}$\tablenotemark{b} & ${\chi_\nu}^2 (\nu)$ \\
\colhead{} & (${\rm arcmin}^2$) & ($10^{22}~{\rm cm}^{-2}$) & \colhead{} & ($10^{-11}~{\rm erg}~{\rm s}^{-1}~{\rm cm}^{-2}$) & \colhead{} & \colhead{} & ($10^{22}~{\rm cm}^{-2}$) & \colhead{} & (keV) & ($10^{-11}~{\rm erg}~{\rm s}^{-1}~{\rm cm}^{-2}$) & \colhead{}
}
\startdata
1 & 108 & 
$0.87\pm0.01$ & $2.39\pm 0.01$& $6.06\pm0.03$ & 1.38 (718) &  & 
$0.77\pm0.01$ & $1.96\pm0.05$ & $9\pm1$ & $5.7\pm0.1$ & 1.04 (717) \\

2 & 77.2 & 
$0.83\pm0.02$ & $2.38\pm0.02$ & $5.22\pm0.05$ & 1.23 (242)  & &
$0.73\pm0.03$ & $2.00\pm0.09$ & $10^{+3}_{-2}$ & $4.9\pm0.2$ &  1.01 (241)\\

3 & 86.6 & 
$0.55\pm0.03$ & $2.62\pm0.04$ & $1.41\pm0.03$ & 1.14 (94) & & 
$0.47\pm0.04$ & $2.2\pm0.2$ & $8^{+7}_{-3}$ & $1.3\pm0.1$ & 1.02 (93) \\

4 & 42.2 &  
$0.54\pm0.04$ & $2.72\pm0.07$ & $0.66\pm0.02$ & 1.38 (44) & & 
$0.44\pm0.07$ & $2.2\pm0.3$ & $6^{+8}_{-2}$ & $0.60^{+0.06}_{-0.05}$ & 1.23 (43)  \\

5 & 116 & 
$0.78\pm0.03$ & $2.39\pm0.04$ & $1.88\pm0.04$ & 1.18 (120) & & 
$0.70\pm0.05$ & $2.1\pm0.2$ & $11^{+13}_{-4}$ & $1.8\pm0.1$ & 1.12 (119) \\

6 & 51.4 & 
$0.88^{+0.06}_{-0.05}$ & $2.43\pm0.06$ & $1.27\pm0.04$ & 1.63 (48) & & 
$0.68\pm0.09$ & $1.7\pm0.3$ & $5^{+3}_{-2}$ & $1.1^{+0.2}_{-0.1}$ & 1.24 (47)  \\ 

7 & 59.6 & 
$0.71\pm0.04$ & $2.38\pm0.05$ & $1.18\pm0.03$ & 1.15 (56) & & 
$0.62\pm0.07$ & $2.0\pm0.2$ & $10^{+14}_{-4}$ & $1.1\pm0.1$ & 1.03 (55)  \\

8 & 72.4 & 
$0.78\pm0.03$ & $2.55\pm0.04$ & $1.91\pm0.04$ & 1.10 (110) & & 
$0.68\pm0.05$ & $2.1\pm0.2$ & $8^{+5}_{-2}$ &  $1.8\pm0.1$ & 0.92 (109) \\

9 & 65.9 & 
$0.81\pm0.03$ & $2.49^{+0.04}_{-0.03}$ & $2.03\pm0.04$ & 1.11 (109)  & &
$0.74^{+0.04}_{-0.05}$ & $2.2\pm0.2$ & $14^{+16}_{-5}$ & $1.9^{+0.2}_{-0.1}$  & 1.03 (108) \\

10 & 54.0 & 
$0.60\pm0.04$ & $2.23\pm0.05$ & $1.49\pm0.03$ & 0.88 (75) & & 
$0.52\pm0.06$ & $1.9\pm0.2$ & $12^{+19}_{-5}$ & $1.4\pm0.1$ & 0.79 (74)
\enddata
\tablenotetext{a}{Errors represent 90\% confidence.}
\tablenotetext{b}{Corrected for absorption.}
\end{deluxetable*}

\subsection{Broad-Band X-Ray Spectral Analysis}\label{subsec:broad}
In this section, we connect the XIS and HXD spectra, which is crucial for 
a quantitative study of the cutoff structure and also for multi-wavelength study described 
in \S\ref{subsec:mws}. In order to study the general characteristics of the emission 
from RX~J1713.7$-$3946, the XIS and HXD spectra obtained in 
\S\ref{subsec:xis_ana}/\S\ref{subsec:hxd_ana} were co-added. 
Then, the summed XIS and HXD spectra were scaled to account for the 
flux from the whole remnant, which makes it easier not only to connect the XIS and 
HXD spectra but also to compare the combined spectrum directly to those of 
other wavelengths in \S\ref{subsec:mws}. 
In scaling the spectra to the whole 
remnant, we assumed the surface brightness of the {\it ASCA} GIS image 
(1--5~keV) shown in Figure~\ref{fig:obs_fov}, the angular response of 
the XIS/XRT system (ARFs), and that of the HXD PIN presented in 
Figure~6 of Paper I. 
The relative normalization factor between XIS and HXD derived from 
Crab observations \citep{ishida06} is included in the scaling. 
Therefore, the scaled 
XIS and HXD can be connected to each other without additional scaling
if no systematic errors are considered.  
However, we estimate that the normalization of the XIS and HXD 
spectra obtained by the procedures above should contain systematic 
errors of 10--20\%. In order to account for the systematic errors, 
we include a constant normalization factor and deal with this factor as a free parameter 
when fitting the XIS+HXD spectrum below. 
In this procedure, we discarded the HXD data from Pointing 4 
since they seem to be contaminated by a nearby hard X-ray source 
as described in \S\ref{subsec:hxd_spatial}.

\begin{figure}
\epsscale{1.0}
\plotone{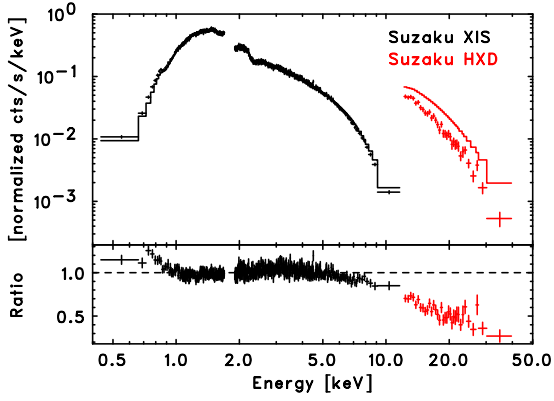}
\caption{The {\it Suzaku} (XIS+HXD) spectrum of RX~J1713.7$-$3946. The model plotted with the data is a power law 
obtained by fitting the XIS data. The parameters are shown in Table~4. The lower panel shows the ratio of the data 
to the model, and clearly reveals a cutoff.}
\label{fig:joint_spec_pl}
\end{figure}

Before the XIS and HXD spectra were jointly analyzed, 
each co-added spectrum was independently fitted using a power-law function. 
Table~4 summarizes the fit results. 
The HXD spectrum gives an acceptable fit.  In contrast, the XIS does not, with a large chi-squared 
value of ${\chi_\nu}^2 = 1.57$ for 711 degrees of freedom and residuals that begin to 
become large at $\sim 6$~keV. 
This fact is clearly seen in Figure~\ref{fig:joint_spec_pl}, where 
the XIS and HXD spectra are plotted together with a power-law function 
which represents the XIS spectrum. 
The spectral steepening begins in the XIS band and continues smoothly into the HXD band. 
This plot strongly suggests that spectral steepening occurs around 10~keV. 
We have already reported such a spectral feature in Paper I for Pointing 0 data. 
The same kind of feature has been revealed in this spectrum averaged over the SNR, 
which suggests that the spectral steepening is common in the entire region. 

We then quantitatively evaluated the spectral steepening by fitting a model with a cutoff 
structure to the combined XIS/HXD spectrum. Below we use our numerically 
calculated synchrotron spectrum which was embedded into the XSPEC package. 
As an electron distribution, we adopt a generalized form, namely, 
\begin{eqnarray}
\frac{dN_e}{dE} \propto E^{-s} \exp  \left[ - \left( \frac{E}{E_0} \right)^\beta \right]\label{eq:e_spec},
\end{eqnarray}
instead of taking a simple 
{\it exponential cutoff} often used in the literature. 
Here, $s$ represents index of electron spectrum and $\beta (> 0)$ determines rapidity of 
high-energy cutoff. 
The photon spectrum can be calculated as 
\begin{eqnarray}
\frac{dN}{d\varepsilon} \propto \varepsilon^{-1} \int F \left(\frac{\varepsilon}{\varepsilon_c}\right) \frac{dN_e}{dE}~dE. 
\end{eqnarray}
Here the function $F(x)$ is defined as 
\begin{eqnarray}
F(x) \equiv x \int^{\infty}_{x} K_{5/3} (\xi)~d\xi,
\end{eqnarray}
where $K_{5/3}$ is the modified Bessel function of 5/3 order. 
When pitch angles are isotropic, the characteristic photon 
energy $\varepsilon_c$ is given as 
\begin{eqnarray}
\varepsilon_c = 0.543 \left( \frac{B}{100~\mu {\rm G}} \right) \left( \frac{E}{10~{\rm TeV}} \right)^2~{\rm keV}. 
\end{eqnarray}
The model consists of four parameters: $s$, 
$\Pi \equiv E_0B^{1/2}$, $\beta$, and the flux normalization. 
Photon spectra were calculated and tabulated for 
reasonable ranges of the four parameters, to be 
used as an XSPEC table model. 
The parameters characterizing the electron energy distribution 
can be obtained directly from the fit to the X-ray data. 

For RX~J1713.7$-$3946, the most probable value for $s$ is 3.0 
rather than 2.0 due to significant synchrotron cooling of electrons 
during the lifetime of the SNR $t_0$ ($\sim 1000~{\rm yr}$). 
The electron spectrum becomes steeper by a factor of 1 ($s \rightarrow s +1$), 
when the injection of electrons is constant and the lifetime 
$t_0$ is smaller than the time scale of synchrotron cooling $t_{\rm sync}$. 
Since $t_{\rm sync}$ can be written as 
\begin{eqnarray}
t_{\rm sync} = 28 \left( \frac{B}{100~\mu{\rm G}} \right)^{-3/2} \left( \frac{\varepsilon}{3~{\rm keV}} \right)^{-1/2}~{\rm yr}, \label{eq:sync_cooling_time}
\end{eqnarray}
the condition above is satisfied in the energy range, 
\begin{eqnarray}
\varepsilon >  \varepsilon_b \equiv 2.3 \left( \frac{B}{100~\mu{\rm G}} \right)^{-3} \left( \frac{t_0}{10^3~{\rm yr}} \right)^{-2}~{\rm eV}. \label{eq:ebreak}
\end{eqnarray}
To explain  the  variability of 
X-ray emission reported on a year timescale from  
compact regions  of the  shell of RX~J1713.7$-$3946,  
\cite{uchi07}  proposed  that  the magnetic field in these 
compact regions is amplified to  1~mG.  The average large-scale 
magnetic field  in the remnant should be significantly lower, but even for $B \simeq 100~{\mu}{\rm G}$ 
the above condition  in the X-ray  domain is safely satisfied. Therefore,  
for a strong shock with a compression ratio of 4.0,
the index of X-ray emitting electrons should be close to $s = 3.0$.

\begin{figure}
\epsscale{1.0}
\plotone{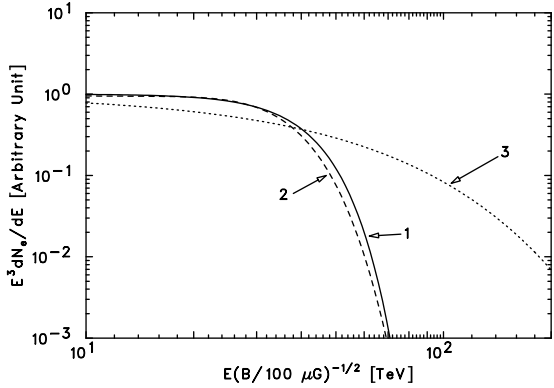}
\caption{Comparison of electron spectra in the high-energy cutoff region. 
Curve 1 is the electron spectrum obtained by fitting the {\it Suzaku} spectrum 
with a synchrotron spectrum assuming the electron distribution given by equation \ref{eq:e_spec} 
when $s$ is fixed to  $3.0$. Curve 2 is the model 
by \cite{zira07}. Both the spectra are drawn by assuming the best-fit parameters of the 
{\it Suzaku} spectrum, which are shown in Table~5. Curve 3 is the {\it exponential cutoff} 
spectrum often assumed in the literature. The cutoff energy is set to the same value as that of 
curve 1.}
\label{fig:cutoff_hikaku}
\end{figure}

We fitted the {\it Suzaku} spectrum with the synchrotron spectrum described above. 
In the fitting procedure, we fixed the 
electron index $s$. Table~5 summarizes the result in the case of $s = 3.0$, which  
corresponds to the most probable case following the discussion above. 
We also present results when $s = 2.0$ for comparison. 
This case can be realized if the magnetic field and/or the age of the remnant are much smaller than we expect, as 
is seen in equation (\ref{eq:ebreak}). 
The spectral fitting with the electron index of $s = 3.0$ yields rather rapid steepening with 
$\beta = 3.4^{+0.7}_{-0.5}$ compared to the conventionally used {\it exponential cutoff}. 
However, one should be careful with the physical interpretation of the fit based on 
equation (\ref{eq:e_spec}). 
Indeed, while the electron spectrum is derived from the broad-band {\it Suzaku} X-ray data 
assuming a specific spectral form given by equation (\ref{eq:e_spec}), the curve 1 in 
Figure~\ref{fig:cutoff_hikaku} has a more general meaning. 
It does not depend on the assumed analytical presentation and, in fact,  relates {\it uniquely} the 
energy spectrum of electrons to the measured X-ray spectrum. 
Indeed, the curve 1 can be presented in different mathematical forms. 
In particular, the electron spectrum shown by curve 1 in Figure~\ref{fig:cutoff_hikaku} 
is quite close to the theoretical prediction for the spectrum of shock 
accelerated electrons in a young SNR \citep{zira07}, 
\begin{eqnarray}
\frac{dN_e}{dE} \propto E^{-3}  \left[ 1+ 0.66 \left( \frac{E}{E_0} \right)^{5/2} \right]^{9/5} \exp \left[ -\left( \frac{E}{E_0} \right)^2 \right]. \label{za07_e_spec}
\end{eqnarray}
This spectrum is derived under the assumption that electrons are accelerated 
by a strong shock in the Bohm diffusion regime and that the energy losses of electrons are 
dominated by synchrotron cooling. 
It is seen that the energy spectrum below the cutoff  is described by a function 
which deviates from a  pure power-law form, 
by an  exponential term with $\beta=2$. 
The latter has a simple physical interpretation and is a result of combination of 
two effects --- acceleration in the Bohm diffusion regime, 
and energy losses in the synchrotron regime. 
The curve 2 shown in Figure~12 is calculated for the parameter 
$\Pi=E_0  B^{1/2} = 201~{\rm TeV}~\mu{\rm G^{1/2}}$. 

The energy spectrum of synchrotron radiation
corresponding to the electron spectrum given by equation (\ref{za07_e_spec}) has a form
\citep{zira07} 
\begin{eqnarray}
\frac{dN}{d\varepsilon} \propto \varepsilon^{-2} \left[ 1+ 0.46 \left( \frac{\varepsilon}{\varepsilon_0}\right)^{0.6} \right]^{2.29} \exp \left[ -\left( \frac{\varepsilon}{\varepsilon_0} \right)^{1/2} \right]. \label{za07_sync_spec}
\end{eqnarray}
This spectrum fits well the broad-band {\it Suzaku} X-ray data for a single parameter 
$\varepsilon_0=0.67 \pm 0.02$ keV. 
Note that the exponential term in the synchrotron 
spectrum is a weak function of energy ($\propto \exp[-(\varepsilon/\varepsilon_0)^{1/2}]$);
therefore the characteristic energy $\varepsilon_0$ only 
formally can be considered as the cutoff energy. 
In fact, for the shape given by equation (\ref{za07_sync_spec}), 
the break in the spectrum starts at much higher energies 
($\varepsilon \sim 10 \varepsilon_0$).

\begin{deluxetable*}{ccccc}
\label{tab:xishxd_all_plfit}
\tablecolumns{5}
\tablewidth{0pc}
\tablecaption{Power-law fitting to the XIS/HXD spectrum of the whole remnant\tablenotemark{a}}
\tablehead{
Detector & $N_{\rm H}$  & $\Gamma$ & Flux\tablenotemark{b} & ${\chi_\nu}^2~(\nu)$ \\ 
\colhead{} & ($10^{22}~{\rm cm}^{-2}$) &  \colhead{} & ($10^{-10}~{\rm erg}~{\rm s}^{-1}~{\rm cm}^{-2}$)  & \colhead{}  
 }
\startdata
XIS (0.4--12~keV) & $0.79\pm0.01$ & $2.39\pm0.01$ & $7.65\pm0.03$ & 1.57 (711) \\
HXD (12--40~keV) & --- & $3.2\pm0.1$ & $1.17\pm0.03$ & 1.16 (67)
\enddata
\tablenotetext{a}{Errors represent 90\% confidence.}
\tablenotetext{b}{Corrected for absorption. The calculated energy band is 1--10~keV and 10--40~keV for 
XIS and HXD, respectively.}
\end{deluxetable*}

\begin{deluxetable*}{cccccc}
\label{tab:xishxd_all_modelfit}
\tabletypesize{\scriptsize}
\tablecolumns{5}
\tablewidth{0pc}
\tablecaption{Fitting to the {\it Suzaku} (XIS+HXD) spectrum\tablenotemark{a}}
\tablehead{
Assumed Function & $N_{\rm H}$ & Constant Factor & Parameter(s) & $F_{1-10~{\rm keV}}$\tablenotemark{a} & ${\chi_\nu}^2~(\nu)$ \\ 
\colhead{} & ($10^{22}~{\rm cm}^{-2}$) &  \colhead{} &  \colhead{} & ($10^{-10}~{\rm erg}~{\rm s}^{-1}~{\rm cm}^{-2}$)  & \colhead{}  
 }
\startdata
Electron Distribution & $0.71\pm0.01$& $1.03\pm0.06$ & $s=3.0$ (fixed), $\beta = 3.4^{+0.7}_{-0.5}$, $\Pi = 402\pm6~{\rm TeV}~\mu{\rm G}^{1/2}$& $7.2\pm0.1$ & 1.11 (778)\\
of Equation (\ref{eq:e_spec})\tablenotemark{c}  & $0.68\pm0.01$ & $1.03^{+0.06}_{-0.05}$ &  $s=2.0$ (fixed), $\beta = 1.5\pm0.2$, $\Pi = 207^{+21}_{-20}~{\rm TeV}~\mu{\rm G}^{1/2}$& $7.1\pm0.1$ & 1.07 (778)\\
 \hline
Z\&A (2007)\tablenotemark{d} & $0.70\pm0.01$ & $1.08\pm0.04$ &  $\varepsilon_0 = 0.67\pm0.02~{\rm keV}$ & $7.2\pm0.1$ & 1.11 (779)
\enddata
\tablenotetext{a}{Errors represent 90\% confidence.}
\tablenotetext{b}{Corrected for absorption.}
\tablenotetext{c}{Details of the calculation of synchrotron spectrum is described in the text.}
\tablenotetext{d}{A model by \cite{zira07}, the synchrotron spectrum of which is shown as equation (\ref{za07_sync_spec}) in this paper}
\end{deluxetable*}

\subsection{Multi-Wavelength Study}\label{subsec:mws}
\subsubsection{Spectral Energy Distribution}\label{subsubsec:sed}
The high quality data of {\it Suzaku}  obtained over two decades
in energy and  combined with  the  TeV gamma ray data, allow 
definite conclusions concerning the origin of multi-TeV parent particles 
based on the comparison of model predictions with observations. In particular 
it is important that the {\it Suzaku}  data provide unambiguous 
information about the shape of the energy spectrum of electrons in the cutoff region. 
 
\begin{figure}
\epsscale{1.0}
\plotone{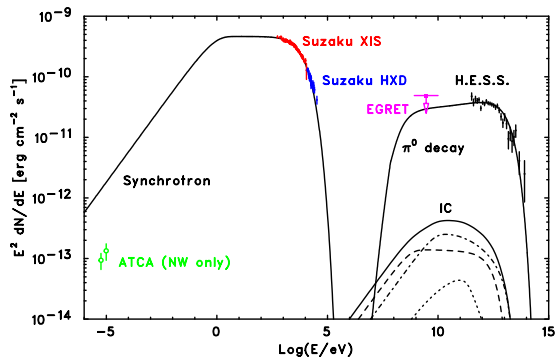}
\caption{Multi-wavelength spectral energy distribution (SED) of RX~J1713.7$-$3946 with 
a hadronic model. The magnetic field is $200~\mu{\rm G}$. The injection index of 
electrons and protons is 2.0, and constant injection during 1000 yr is assumed. For the 
IC flux, the contribution of each photon field is shown. The dashed line, the dash-dotted line, and 
the dotted line are IC flux of CMB, infrared, and optical photons, respectively. 
By taking 1~kpc as the distance of the SNR, the total energy of electrons is 
$W_e = 3.1 \times 10^{46}~{\rm erg}$. The total proton energy is 
$W_p = 2.7 \times 10^{50}~(n/1~{\rm cm}^{-3})^{-1}~{\rm erg}$, 
where $n$ is the ambient matter density.}
\label{fig:SED}
\end{figure}

Figure \ref{fig:SED} shows the Spectral Energy Distribution (SED), $E^2 dN/dE$, 
of  RX~J1713.7$-$3946 from radio to TeV energies.  
The X-ray data correspond to the entire remnant; they  
are reconstructed assuming the best-fit model given by  
equation (\ref{za07_sync_spec}) corrected for the interstellar absorption with 
$N_{\rm H} = 0.70 \times 10^{22}~{\rm cm}^{-2}$. 
The points in the TeV gamma-ray range  corresponding  to the 
fluxes of the whole remnant, 
are from the latest  report of the H.E.S.S. collaboration \citep{aha07}.    
The EGRET upper limit is taken from \cite{aha06}; it is obtained  
through  modeling and subtracting the  resolved  EGRET source 3EG~1714$-$3857 \citep{hartman99}. 
The two data points in the radio band  are shown as measured  with the  ATCA telescope \citep{laz04}.
But it should be noted that  these  fluxes  are  detected only from the NW rim 
region,  therefore they  should be treated as lower limits when  compared to 
model predictions for the whole remnant.  The  recent estimates based on 
observations with  ATCA and  the 30-m radiotelescope  IRA show that the flux of the entire remnant 
should be  a factor of  few larger than  the flux of the NW rim 
(F. Acero et al. 2008, in preparation; G. Dubner 2008, private communications). 

In Figure \ref{fig:SED} we show  theoretical fluxes of broad-band electromagnetic 
radiation  produced by both accelerated electrons and protons.  
The results depend strongly on the strength of the average magnetic field $B$.
Although  the magnetic field in  compact filaments can be as large as 
$\sim 1~{\rm mG}$ \citep{uchi07}, the field in less
bright and more diffuse regions where the bulk of  the synchrotron X-ray 
emission is produced,  should be significantly weaker.  Yet, a moderately 
large field exceeding $100~\mu{\rm G}$ is  
required to explain the X-ray flux ratio between the diffuse and compact zones. 
The integrated X-ray flux from the diffuse zone (zone 1) is larger by a factor of 
$\sim 4$ than that from the compact zone (zone 2): $F_1 \sim 4 F_2$ \citep{uchi03}. 
Using the ratio of the emission volumes estimated from the X-ray image 
($V_1 \sim 1000 V_2$), the number density of X-ray emitting electrons, 
$n \propto F/(VB^2)$, in the zone 2 relative to the zone 1 is estimated as 
$n_2  \sim 250(B_1/B_2)^2 n_1$. For $B \gtrsim 1~{\rm mG}$, and assuming  
$n_2  \geq  n_1$, one finds  $B_1 \geq  100~\mu{\rm G}$. 

In Figure~\ref{fig:SED} the synchrotron, inverse Compton (IC), and $\pi^0$-decay fluxes  
are calculated  assuming  strong uniform magnetic field of strength $B=200~\mu{\rm G}$. 
The calculations are performed with constant injection of electrons and protons over 
the last 1000~yr.  The injection spectrum for both electrons and protons is assumed to be a power-law 
with an index $s=2.0$.  For calculations of IC fluxes  we used the  diffuse photon  
fields  proposed  by \cite{porter06}, including  2.7~K CMB, as well as 
interstellar radiation consisting of two, optical (starlight) and infrared (dust) emission components.
      
The best fits for the synchrotron radiation and $\pi^0$-decay gamma rays are achieved for 
cutoffs in the energy distributions of electrons and protons at 28.4~TeV and 130~TeV, respectively.
The spectra of $\pi^0$-decay gamma rays are calculated using the  analytical presentations by \cite{kelner06}.
Since the highest energy tail of the observed  gamma-ray spectrum is best fitted with
$\exp [-(\varepsilon/\varepsilon_0)^{\beta_\gamma}]$ with $\beta_\gamma  \approx 0.5$ \citep{aha07}, 
the proton spectrum in the cutoff region should have an exponential behavior,
$\beta_p \approx 2 \times \beta_\gamma =1$ (see \citealt{kelner06}). 

The electrons suffer synchrotron losses, thus the electron spectrum
above a certain energy becomes  steeper   ($E^{-s}  \rightarrow E^{-(s+1)}$). 
The position of the break in the electron spectrum appears at 
\begin{eqnarray}
E_b = 1.25 \left( \frac{B}{100~\mu{\rm G}} \right)^{-2} 
\left( \frac{t_0}{10^3~{\rm yr}} \right)^{-1}~{\rm TeV}\label{eq:ebreak_electron}. 
\end{eqnarray}
For a magnetic field $B=200 \mu \rm G$ and age of the source  $T \leq 1000$~yr,
the spectral  break in the synchrotron spectrum  corresponding to the transition of 
the electrons spectrum from uncooled to cooled regime,  appears around 1 eV 
(see Figure~\ref{fig:SED}).   Thus, for young SNRs  the detection of the synchrotron 
break  at optical/infrared wavelengths would be an additional argument in favor of strong 
magnetic field.  The cooling  break in the electron spectrum is reflected also in the 
gamma-ray band. Namely, for $B=200 \mu \rm G$ the corresponding signature 
in the IC gamma-ray spectrum appears around 10 GeV. 
Unfortunately (see Figure~\ref{fig:SED}),  for such a strong magnetic field, the IC 
component is suppressed, and falls well below the sensitivity of gamma-ray detectors in this 
energy band, including {\it GLAST}.  At the presence of such a strong magnetic field, 
the only viable mechanism which can produce  TeV gamma-rays at the flux level 
detected by H.E.S.S.  is related to interactions of  ultrarelativistic protons with ambient gas
through production and decay of  $\pi^0$-mesons.  To explain  the  detected flux of gamma-rays
protons should be accelerated to energies  well beyond 100 TeV, and  the parameter
$A=(W_{\rm p}/10^{50} \rm erg) (n/1 \ cm^{-3})  (d/1 \rm kpc)^{-2}$ should be between 1.5--3, depending 
on the spectrum of protons. 

\begin{figure}
\epsscale{1.0}
\plotone{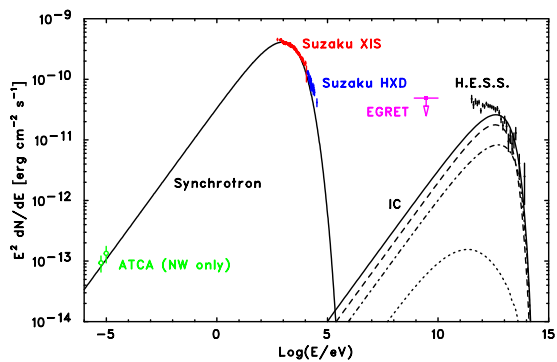}
\caption{The SED of RX~J1713.7$-$3946 with a leptonic model. The injection index of 
electrons is 2.0. The calculation is for magnetic field strength of $14~\mu{\rm G}$. 
The total energy of electrons is $W_e = 1.4 \times 10^{47}~{\rm erg}$
The line styles for the IC spectra are the same as those in Figure~\ref{fig:SED}. 
}
\label{fig:SED_leptonic}
\end{figure}

\begin{figure}
\epsscale{1.0}
\plotone{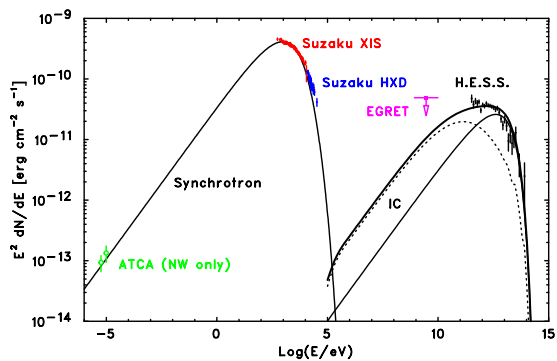}
\caption{Same as \ref{fig:SED_leptonic} but additional optical 
radiation is included as seed photons of IC scattering to fit the 
H.E.S.S. spectrum. 
The IC spectrum with the thin solid line and the dotted line correspond 
to the contribution of the interstellar radiation assumed in Figure~\ref{fig:SED_leptonic} 
and the additional optical radiation, respectively. 
The thick solid line indicates the sum of the two. 
The energy density of the additional optical 
radiation is $140~{\rm eV}~{\rm cm}^{-3}$, which 
is about two order of magnitude larger than 
the estimate by \cite{porter06}.}
\label{fig:SED_leptonic_opt}
\end{figure}

In order to increase the IC flux to the level of the observed 
TeV gamma-ray flux, the magnetic field should  be reduced down to 
 $10~\mu{\rm G}$ and $15~\mu{\rm G}$.  For the given  magnetic field, 
 the high quality X-ray data of {\it Suzaku}  obtained over two energy decades 
allow derivation of the electron spectrum with high accuracy within the interval 
covering one energy decade:  from $\sim 50(B/10~\mu{\rm G})^{-1/2}~{\rm TeV}$ to 
$\sim ~500(B/10~\mu{\rm G})^{-1/2}~{\rm TeV}$. 
This allows us to calculate the spectrum and absolute flux of IC gamma rays above 
a few TeV without any model assumptions, as long as the main target for the IC 
gamma-ray production remains the 2.7 K CMB. The contribution of the diffuse optical/infrared
radiation fields  generally is less, however the optical photons may provide enhanced TeV 
emission at low, sub-TeV energies.   For  calculations of the IC spectrum we used  the 
interstellar radiation model  of  \cite{porter06}, who  proposed  significantly 
larger  flux  of  optical and infrared components compared to the generally accepted flux. 
However, the results presented in Figure~\ref{fig:SED_leptonic} show that even 
this  high diffuse optical  and infrared radiation  fails to account for the 
observed gamma-ray flux below a few TeV. 
In order to fill this gap one needs to assume an unreasonably large density of optical radiation. 
This is demonstrated in Figure~\ref{fig:SED_leptonic_opt}, where an agreement of IC 
calculations with the  reported gamma-ray fluxes is achieved assuming an additional, although 
in our view quite  unrealistic, component of optical radiation with a density of $140~{\rm eV}~{\rm cm}^{-3}$. 

It should be noted that the problem of explanation of low energy gamma rays in Figure~\ref{fig:SED_leptonic}  is related to the large  energy of the break in the electron spectrum given by 
equation~(\ref{eq:ebreak_electron}),  and correspondingly to the  position of 
the Compton peak which appears above 1~TeV in  the spectral energy distribution of gamma rays. 
Thus,  the reduction of the break energy down to 200~GeV could in principle solve the 
problem. However, since the magnetic field in this model cannot exceed $15~\mu{\rm G}$, 
the only way to shift the Compton peak to sub-TeV energies is to assume that the 
supernova remnant is older than  $10^4$~yr. 

\begin{figure}
\epsscale{1.0}
\plotone{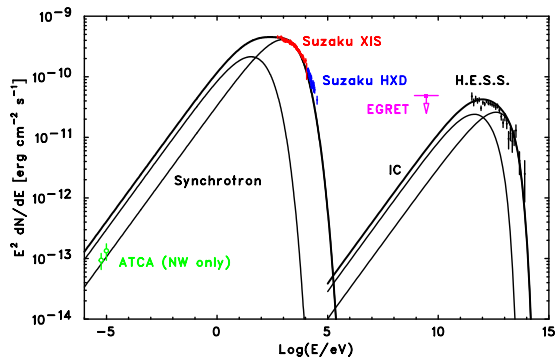}
\caption{Same as \ref{fig:SED_leptonic} but another electron population 
is added. The thin solid lines are radiation from each electron population 
and the thick solid line is the sum of the contributions from the two electron 
populations. The total energy of the second electron 
population is $W_e = 3.4 \times 10^{47}~{\rm erg}$}
\label{fig:SED_leptonic_2ele}
\end{figure}

An alternative  solution for explanation of  gamma-ray data within the IC models 
is to assume a higher density of relativistic electrons responsible for 
$\leq 1$ TeV gamma rays,  i.e. to postulate  a second, 
low-energy electron component in the shell. 
Formally this assumption does not {\it a priori} contradict the observations since  
the additional electrons are required  to be present below  20~TeV,  
i.e. in the energy range  which is not constrained  by  {\it Suzaku} observations
 (for production of the lowest energy X-rays detected by {\it Suzaku} in a 
magnetic field of $15~\mu{\rm G}$ the electron energy must exceed 40~TeV).
The results calculated under such assumption are shown in Figure~\ref{fig:SED_leptonic_2ele}. 
The second electron component is assumed to have a power-law 
injection spectrum with the same index  as the first (main) component, $s=2.0$, 
but with a high energy cutoff around 10~TeV.  The latter is  required  to prevent  the 
conflict with the observed gamma-ray spectrum above 1~TeV.


\subsubsection{Morphology}
A comparison between the X-ray and the TeV gamma-ray morphology 
is expected to provide information 
about the acceleration/emission processes. As is already shown in 
Figure~\ref{fig:xis_hess_image}, the {\it Suzaku} XIS image revealed 
a significant correlation with that of H.E.S.S. telescopes not only in the 
bright structures but also in the dim regions of the remnant. In the following, 
the two images are compared with each other  on a more quantitative basis. 

\begin{figure}
\epsscale{1.0}
\plotone{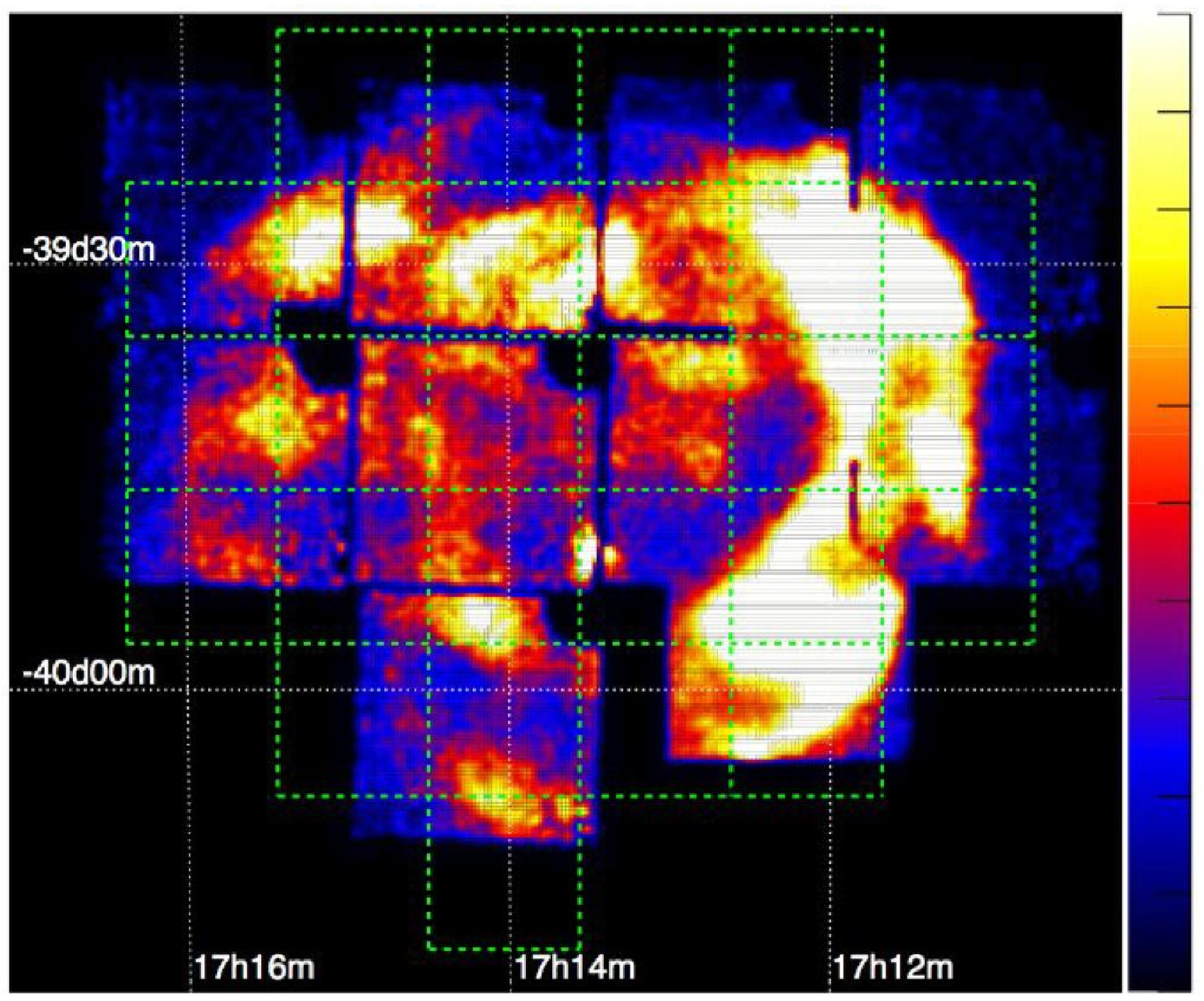}
\plotone{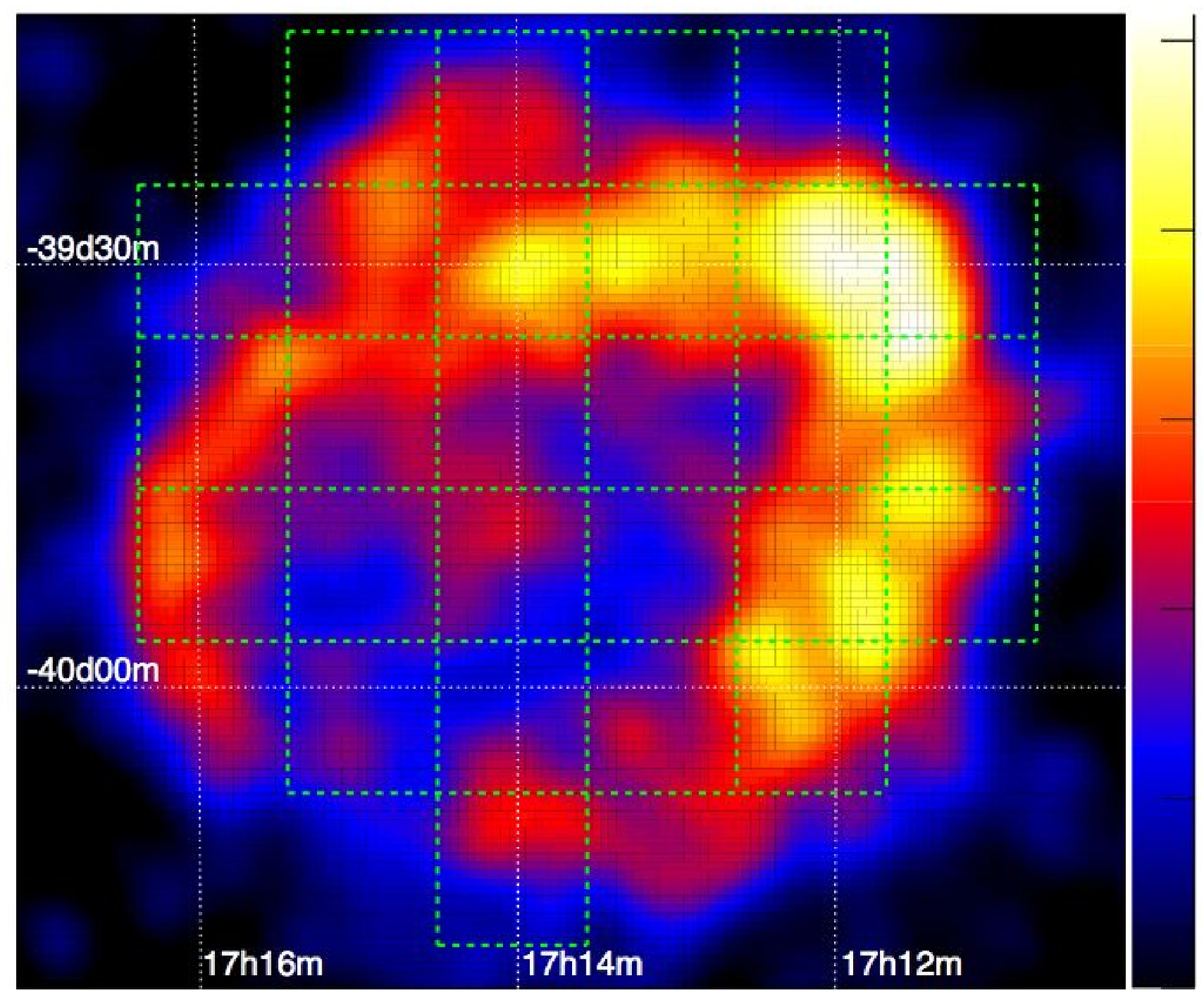}
\caption{The regions for comparison of the X-ray and the TeV gamma-ray morphologies overlaid 
on the XIS image (2--5~keV) (upper panel) and the H.E.S.S. image \citep{aha07} (lower panel).  
The size of each square region 
is $10^{\prime}.8 \times 10^{\prime}.8$. The results of the comparison are shown in 
Figure~\ref{fig:plot_keV_vs_TeV} and Figure~\ref{fig:map_keV-TeV}.}
\label{fig:mosaic_image_sqreg}
\end{figure}

It should be taken into account that  the X-ray morphology can be affected by the 
spatial distribution of absorption column density, $N_{\rm H}$. In order to avoid 
the effect, here we disregard  the energy band lower than  
1~keV, and use two energy intervals, 2--5~keV and 5--10~keV,  for  comparison. 
For the $N_{\rm H}$ variation as shown in Table~3, the count rate 
can vary by 5\% and 0.7\% in the 2--5~keV band and 5--10~keV band, respectively. 
One should also take into account the difference of point spread functions between 
{\it Suzaku} XIS and H.E.S.S.  when comparing the two images. To prevent  
this, we compare  surface brightness for each square region 
with a size of $10^{\prime}.8 \times 10^{\prime}.8$, which is larger than 
the point spread functions of either observatory. The regions are indicated 
with green dashed lines in Figure~\ref{fig:mosaic_image_sqreg}. 

\begin{figure}
\epsscale{1.0}
\plotone{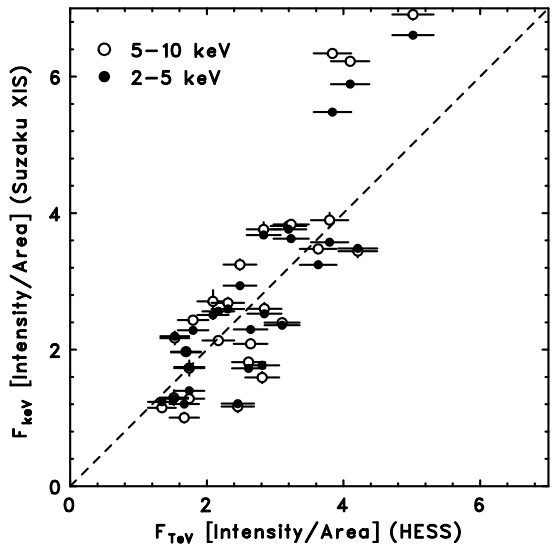}
\caption{The scatter plot of X-ray surface brightness from the XIS and the gamma rays from H.E.S.S. for each square region shown 
in Figure~\ref{fig:mosaic_image_sqreg}. Each data set is scaled for direct comparison.}
\label{fig:plot_keV_vs_TeV}
\end{figure}

\begin{figure}
\epsscale{1.0}
\plotone{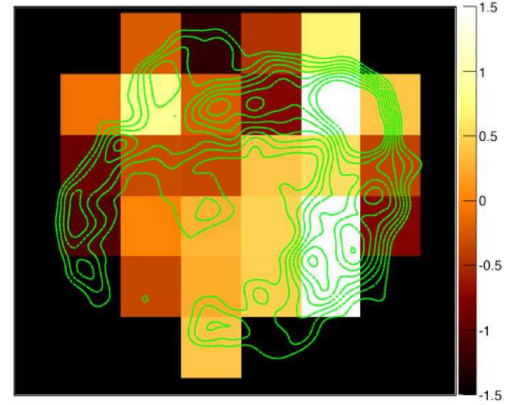}
\caption{The map of $F_{\rm keV} - F_{\rm TeV}$, both of which are plotted in Figure~\ref{fig:plot_keV_vs_TeV}. 
The values of  $F_{\rm keV} $ are those for the energy range of 2--5~keV. The overlaid contours are the H.E.S.S. image 
taken from \cite{aha07}.}
\label{fig:map_keV-TeV}
\end{figure}

Figure~\ref{fig:plot_keV_vs_TeV} shows a scatter plot between the
{\it Suzaku} XIS count rate ($F_{\rm keV}$) and that of H.E.S.S. ($F_{\rm TeV}$). 
As seen in this plot, the X-ray count maps correlate strongly with the 
gamma-ray count map. The correlation coefficients are calculated to be 0.85 
and 0.83, for 2--5~keV band and 5--10~keV band, respectively. 
It is worth noting that there are some deviations (X-ray intensity excesses)  in 
the bright regions. Figure~\ref{fig:map_keV-TeV} shows a map of 
$F_{\rm keV} - F_{\rm TeV}$ (for 2--5~keV band) overlaid with 
the H.E.S.S. contours, in which one can see that the X-ray excesses are present 
along the NW and SW rims.

\section{DISCUSSION}
\subsection{Cutoff in the Synchrotron Spectrum}
We conducted a series of {\it Suzaku} observations 
which covers about two-thirds of the surface of SNR RX~J1713.7$-$3946.  
Through the data analysis, we successfully detected 
signals up to $\sim 40$~keV from each of the pointings. 
The HXD spectra above 10~keV are significantly 
steeper than those obtained from the XIS below 
10~keV, suggesting that a spectral cutoff 
is common throughout the remnant. By combining the XIS 
and HXD spectra, we obtained a wide-band spectrum with 
high statistics, which clearly shows a cutoff around 
10~keV. 

Taking advantage of the high photon statistics, we performed a detailed study of the cutoff shape and 
compared it with a recent theoretical prediction by \cite{zira07}. 
A sharp cutoff of the accelerated electron spectrum is needed to reproduce the cutoff shape in the synchrotron spectrum 
detected with {\it Suzaku}.  The spectrum of electrons derived from  
{\it Suzaku} data is in good agreement with the analytical model of  \cite{zira07}.

The cutoff energy in the spectrum of synchrotron radiation contains 
an important information about the efficiency of diffusive shock acceleration. 
For  acceleration in the Bohm diffusion regime and when energy losses of electrons are dominated by synchrotron cooling,  the cutoff energy,  $\varepsilon_0$ in 
equation (\ref{za07_sync_spec})  is expressed as \citep{zira07}  
\begin{eqnarray}
\varepsilon_0 = 0.55 \left( \frac{v_s}{3000~{\rm km}~{\rm s}^{-1}} \right)^2 \eta^{-1}~{\rm keV}, 
\end{eqnarray}
where $v_s$ is the shock speed and $\eta~(\ge1)$ is the so-called ``gyrofactor''. 
The case of  $\eta = 1$ corresponds to 
the  ``Bohm limit'', and implies  high level of turbulence $\delta B \sim B$. The  
{\it Suzaku} spectrum is characterized  by the  
best-fit  parameter $\varepsilon_0 = 0.67~{\rm keV}$ which  
gives $v_s =  3300 \eta^{1/2}~{\rm km}~{\rm s}^{-1}$. 
Here, we assume that the shock speed $v_s$ is uniform throughout the remnant, which 
is supported by the fact that the outer boundary of the X-ray morphology 
is nearly circular. The 
upper-limit of the shock speed $v_s \le 4500~{\rm km}~{\rm s}^{-1}$ 
derived from  the   {\it Chandra} data 
\citep{uchi07}, results in   $\eta \le 1.8$. 
This is a strong  evidence of 
acceleration of electrons in the regime close to the Bohm limit. 
Note that a similar result was obtained for the SW rim of the remnant in \cite{uchi07}. 
Here we confirm this conclusion for a larger area of the remnant with higher statistics. 

\subsection{Multi-wavelength Spectrum}

While there is little doubt in the  synchrotron origin of broad-band X-ray emission 
measured by {\it Suzaku},  the X-ray spectrum alone does not give preference to the
strength of the magnetic field in the region of production of synchrotron radiation. Formally,
the field can be as small as $10~\mu{\rm  G}$ and as large as $100~\mu{\rm G}$. Meanwhile,  
the strength of the magnetic field has dramatic impact on the origin of TeV gamma-rays. 
The so-called leptonic or inverse Compton models require magnetic field between
$10~\mu{\rm G}$  and $15~\mu{\rm G}$.  Even so, it is difficult to achieve, at least 
within a simple one-zone model,  a  satisfactory explanation of both X-ray and TeV
gamma-ray spectral features, unless we invoke an extremely high diffuse 
radiation field of optical  photons  to enhance the IC  
gamma-radiation below 1~TeV (see  Figure~\ref{fig:SED_leptonic_opt}).    
A more realistic approach  for  explanation of  the broad-band TeV
gamma-ray spectrum  within IC models  can be realized under the 
assumption of existence an additional,  low-energy electron component in the shell 
(see Figure~\ref{fig:SED_leptonic_2ele}). 
Even so, the most serious  problem for IC models remains the requirement of low magnetic field in the
gamma-ray production region,  in contrast to large magnetic field 
required to explain the fast  variability of X-ray emission on small scales.
Formally, one may assume that gamma-rays are mainly produced in "voids", i.e.
in regions  with very low magnetic field. This would imply quite inhomogeneous 
distribution of the magnetic field in the shell. One the other hand,  
the  observed strong X-ray and TeV correlation within the IC models can be 
explained  only in the case of homogeneous distribution of  magnetic field.  

\begin{figure}
\epsscale{1.0}
\plotone{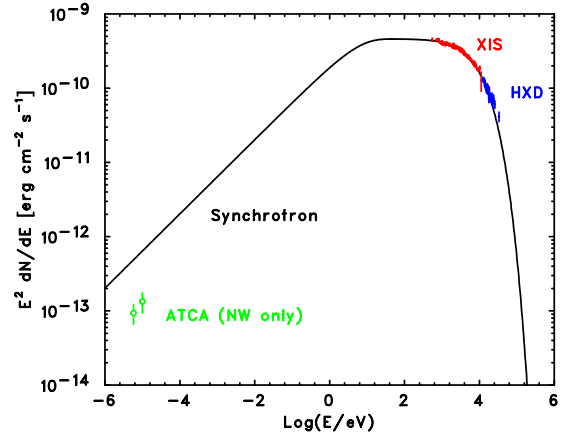}
\caption{The SED of RX~J1713.7$-$3946 from radio to X-ray with a synchrotron model curve 
calculated by assuming $B = 100~\mu{\rm G}$. 
}
\label{fig:SED_100uG}
\end{figure}

The large-scale magnetic fields on parsec scales with an average  strength larger than 
$\geq 15~\mu{\rm G}$  make the IC gamma-ray production inefficient, and thus 
give preference to the so-called hadronic models of 
gamma-rays produced at interactions of accelerated protons with the 
ambient gas via production and decay of secondary $\pi^0$-mesons.
What concerns X-rays, they are produced,  as in leptonic models,  
by synchrotron radiation  of directly accelerated electrons.
This is demonstrated in  Figure~\ref{fig:SED} for  very strong magnetic field, 
$B=200~\mu{\rm G}$. Note that  while comparing the model predictions with 
measurements in the radio band, one should take into account that 
the radio points shown correspond to measurements of NW rim, while 
the X-ray and gamma-ray points are for the entire remnant.  
If the ratio of the radio flux from the NW rim to that from the whole remnant is 
not much different  from  the corresponding ratio in X-rays, 
the flux from the whole SNR should be significantly larger. This would  
reduce the difference between the measurements and predictions.  
In any case,  the radio flux can be significantly 
reduced assuming somewhat smaller magnetic field or 
harder electron spectrum.  Indeed, in Figure \ref{fig:SED_100uG} we 
show model calculations performed for a magnetic field  $B=100~\mu{\rm G}$.
While the synchrotron X-ray flux is described perfectly as before (in Figure~\ref{fig:SED}),
the radio flux is  by a factor of four lower; at 1.4~GHz it is 34~Jy, which is 
close  to the latest estimates of radio flux from the whole remnant based 
on observations with ATCA and the 30-m radio telescopes of IRA 
(F. Acero et al. 2008, in preparation; G. Dubner 2008, private communications).

\begin{figure}
\epsscale{1.0}
\plotone{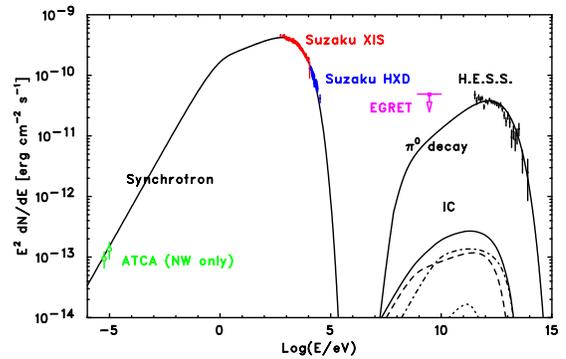}
\caption{The SED of RX~J1713.7$-$3946 with a hadronic model 
when the index of the electron/proton spectrum is 1.7. 
The magnetic field is $200~\mu{\rm G}$. 
The line styles for the IC spectra are the same as those in Figure~\ref{fig:SED}
the total energy of electrons is 
$W_e = 6.0 \times 10^{45}~{\rm erg}$. The total proton energy is 
$W_p = 1.6 \times 10^{50}~(n/1~{\rm cm}^{-3})^{-1}~{\rm erg}$. 
}
\label{fig:SED_17}
\end{figure}

The  radio flux  can be suppressed  even for the ambient 
field larger than $100~\mu {\rm G}$,  provided that the   
electron injection spectrum is harder than $E^{-2}$. 
Figure~\ref{fig:SED_17} demonstrates this  possibility, where we assume an 
electron/proton index of $s = 1.7$  which corresponds to a compression ratio of 
$\sigma = 5.3$. Note that $\sigma$ can exceed the adiabatic upper limit of 4 
as described by \cite{bere06}.  Note that the value of $s = 1.7$ is consistent with 
the conclusion of  \cite{villa07}  based on semi-analytical 
derivation of the  parent proton  spectrum  from the H.E.S.S. data. 
In model calculations shown  in Figure~\ref{fig:SED_17},
the spectrum of protons requires  an ``early''  exponential cutoff  
at $E_{p0} = 25.0~{\rm TeV}$.  Note that formally the spectral index $s=1.7$ 
implies  shock acceleration in non-linear regime which in fact  predicts 
some deviation from pure power-law distribution  of accelerated particles 
(see e.g. \citealt{Ellison07,bere06}). This  would lead to further reduction of the radio flux. 
 
The convection of low energy electrons could be another reason for 
low radio flux. Note that the escape 
of electrons through convection has strong impact only on low energy 
electrons; because of fast synchrotron cooling,  the effect of escape 
is negligible for multi-TeV electrons. A significant quantity of low-energy 
electrons can escape from the  shell of the SNR before emitting radio photons. 
Therefore, the radio flux can be reduced while the X-ray flux will remain unchanged. 

For calculations  shown in Figure~\ref{fig:SED}, the 
total energy of electrons is estimated as 
$W_e =  3.1 \times 10^{46}~(d/1~{\rm kpc})^2~{\rm erg}$,  and the energy for protons as
$W_p = 2.7 \times 10^{50}~(n/1~{\rm cm}^{-3})^{-1}(d/1~{\rm kpc})^2~{\rm erg}$. 
In the case of harder energy spectra with power-law index $s=1.7$ 
corresponding to Figure~\ref{fig:SED_17},  one has 
$W_e =  6.0 \times 10^{45}~(d/1~{\rm kpc})^2~{\rm erg}$,  and 
$W_p = 1.6  \times 10^{50}~(n/1~{\rm cm}^{-3})^{-1}(d/1~{\rm kpc})^2~{\rm erg}$, 
respectively.  The proton/electron ratio in either case is very small, 
$K_{ep} \leq 10^{-4} (n/1~{\rm cm}^{-3})$. This value is significantly 
smaller  than that for directly observed local cosmic rays 
($K_{ep} \sim 0.01$),  unless a large ambient matter density 
of $n \sim 100~{\rm cm}^{-3}$ is assumed. 
\cite{katz08} and \cite{butt08} argued that the hadronic scenario for this SNR has difficulties 
because the $K_{ep}$ value should be consistent with local cosmic rays 
and other SNRs. 
However, the $K_{ep}$ value of one SNR 
at a fixed age does not necessarily need to agree with the local cosmic ray 
value. The low-energy electrons are likely produced in later stages of SNR 
evolution, when the $K_{ep}$ value can be different from the present 
value. A comparison with other SNRs should be performed with care, as well. 

Cutoff energy in the gamma-ray spectrum should give an important hint 
whether SNRs are sources of cosmic rays below the {\it knee}, if gamma rays 
observed by H.E.S.S. have hadronic origins. \cite{plaga08} argued 
that the cutoff energy of the H.E.S.S. spectrum of RX~J1713$-$3946 is 
around 18~TeV, which can be translated to proton energy more than 10 times 
below the energy of the {\it knee}. 
Indeed our multi-wavelength study requires a cutoff in the proton 
spectrum around 100~TeV or even less for hard 
acceleration spectra of protons. However, one should take into account that even 
in the case of effective acceleration the highest energy protons beyond 100 TeV 
escape the source in a quite short time scales, and hence do not contribute to the
gamma-ray production at the present epoch \citep{ptuskin05, gabici07}. 

A  unique feature of RX~J 1713.7-3946 is the lack of thermal X-ray emission. 
Recently, \cite{katz08} and \cite{butt08}  interpreted this fact as an argument against the hadronic model 
for TeV gamma rays. 
Generally it is true that plasma in young supernova remnants is heated to high temperatures observed via 
thermal X-ray emission of hot electrons. However, one should take into account that we deal with a unique 
object, and the lack of thermal X-ray emission cannot {\it a priori} be invoked as an argument against the hadronic origin  of the observed TeV gamma rays. 

It is important to note that in SNR shocks the formation of high plasma temperatures with 
$kT_i=3/16~m_i v_s^2$ is relevant only to protons (ions), and that a high ion 
temperature does not automatically (from first principles) mean a high electron temperature. In fact, 
the only known heating process of thermal electrons is Coulomb collisions between electrons and 
protons (ions), which, however, has too long time scale to establish electron-proton equipartition.
On the other hand, we do know from X-ray observations that the electrons 
in young SNRs are heated to keV temperatures.  This can be explained by assuming that a hypothetical 
mechanism, most likely related to the energy exchange through excited plasma waves, is responsible 
for effective electron heating in SNRs. As long as the nature of this mechanism in 
collisionless shocks remains unknown, one cannot predict, even qualitatively, the specifics of its operation on a source by source basis. 

We indeed deal with two interesting facts. First, many young SNRs, like Tycho and Cassiopeia A, 
with intense thermal X-ray emission and intense nonthermal radio emission emit little (or do not emit at all) 
TeV gamma rays. On the other hand, RX~J1713.7$-$3946, with lack of (or rather very low)  thermal X-ray 
emission and with relatively weak nonthermal radio emission, is a source of powerful TeV radiation.  
These two facts can be treated as a hint for low efficiency of establishing equipartition in the thermal 
plasma in very effective TeV particle accelerators like RX~J1713.7$-$3946. It is interesting to note in this 
regard, that such a tendency is found also for  
another effective TeV accelerator --- SN 1006  \citep{vink03}. 
Whether the reduction of the exchange rate between different particle species in thermal plasma 
has a link to the particle acceleration in high Mach number shocks, as proposed by \cite{vink03}, 
is a very interesting question to be explored in future deep theoretical and phenomenological studies. 
In this regard, RX~J1713.7$-$3946 can serve as a key ``template'' source for such studies.

\cite{hughes00} also discussed low electron temperature based on {\it Chandra} 
observations of a young SNR in the Small Magellanic Could, 1E~0102.2$-$7219. 
They measured a blast-wave velocity of $\sim 6000~{\rm km}~{\rm s}^{-1}$ from the 
expansion rate and predicted the electron temperature of $kT_e > 2.5~{\rm keV}$ 
by considering Coulomb heating. 
However, the electron temperature derived from their spectral analysis is 0.5~keV, which 
is far below the prediction. According to their discussion, not only electron heating 
but also ion heating is suppressed and substantial fraction of energy may be going into 
cosmic-ray production due to the non-linear effects in the shock. 

The nonlinear shock acceleration in this object can convert a significant, 
up to  $f  \sim  0.5$  fraction of the  kinetic energy of explosion into relativistic particles.
Correspondingly the fraction of  available energy which goes to the 
heating of the ambient  plasma will be reduced  $1-f \sim 0.5$. Yet, conservative estimates 
show that plasma in RX~J1713.7$-$3946 can be heated to quite high temperatures even 
the heating of  electrons and  protons proceeds only through the Coulomb exchange. 
This question recently has studied by \cite{Ellison07} for a standard SNR of age $t_{\rm SNR}=500$~yr
and  energy $E_{\rm SN}=10^{51}$~erg. In particular, it has been shown that in the case 
of effective diffusive shock acceleration and the 
plasma density $n=0.1~{\rm cm}^{-3}$, the ratio of synchrotron luminosity 
to thermal (bremsstrahlung) luminosity can be as large as 100. This implies that 
in the case of RX~J1713.7$-$3946,  from which thermal X-ray emission is not observed, 
the plasma density cannot  significantly exceed $0.1~{\rm cm}^{-3}$ (the luminosity of 
thermal bremsstrahlung is proportional to $n^2$).  The low density of order of 
$0.1~{\rm cm}^{-3}$  reduces the parameter space for hadronic models but 
does not exclude it. Indeed, as mentioned above,  the TeV 
gamma-ray flux  of RX~J1713.7$-$3946 can be explained  by interactions of 
protons if the parameter $A=(W_p/10^{50}~{\rm erg}) (n/1~{\rm cm}^{-3})  (d/1~{\rm kpc})^{-2}$
exceeds 1.5   to 3, depending on the spectrum of protons. Assuming that  more than 
$30\%$ of the explosion energy of this SNR is released in accelerated protons, and 
that  the plasma in the  gamma-ray production region is compressed by a factor of few,
we find that the  located of the source at a distance of about 1 kpc would marginally
support the hadronic model.  While  closer location of the source would make 
the model requirements quite viable and flexible, the location of  the source beyond 1~kpc
hardly can be accommodated within a standard shock acceleration scenario.  
If the SNR is closer, the distance of $d \gtrsim 0.5~{\rm kpc}$ seems reasonable 
considering the upper limit on the shock speed of $4500(d/1~{\rm kpc})~{\rm km}~{\rm s}^{-1}$ 
by \cite{uchi07}. 

Finally we should mention the model  suggested by \cite{Malkov05} which can naturally 
explain both the low synchrotron flux at radio frequencies and lack of thermal X-ray emission  
of  RX~J1713.7$-$3946. The standard scenarios of gamma-ray production in SNRs assume that 
radiation is produced in downstream where the density of both  relativistic particles and thermal 
plasma is higher than in upstream.  However,  in the  cases when the shock 
is expanding into a low-density  wind bubble and approaching  cold dense material, 
e.g. swept-up shell or  surrounding molecular clouds, the gamma-radiation 
is contributed predominantly from upstream. While the energy distribution 
of accelerated particles downstream is coordinate-independent in both linear 
and nonlinear regimes, the  particle distribution upstream  is coordinate-dependent.   
Because of energy-dependent diffusion coefficient, 
the high-energy particles diffuse ahead of low-energy  particles, thus a dense material 
adjacent upstream will ``see'' relativistic particles (protons and electrons)
with low-energy exponential cutoff,  $E_{\rm min}$,  which depends on the location of the  dense regions. 
This implies that the effective production of TeV  gamma-rays (from {\it p-p} interactions) 
and X-rays (from synchrotron radiation of TeV electrons) will be not accompanied by low energy 
(GeV) gamma-rays and synchrotron radio emission. Obviously, this model is not constraint by 
lack of thermal emission.  

\subsection{Morphology}
In addition to the spectral information, the comparison of  
X-ray and TeV gamma-ray images presented in Figure~\ref{fig:plot_keV_vs_TeV} helps us 
to draw the physical picture of RX~J1713.7$-$3946. 
Let us first discuss the tight correlation observed in most parts of the remnant.
Within the hypothesis of hadronic origin of gamma-rays,   
the gamma-ray flux is proportional to the number densities of the ambient matter and 
relativistic protons, while the X-ray flux is proportional to number density of electrons. 
If the matter distribution significantly varies throughout the SNR, we need fine 
parameter tuning among the matter distribution, the electron injection rate, and the 
proton injection rate, in order to produce the tight correlation.  
Therefore, more natural explanation is that the matter density is uniform and 
the injection rate of the electrons and that of the protons are proportional to each other. 

\begin{figure}
\epsscale{1.0}
\plotone{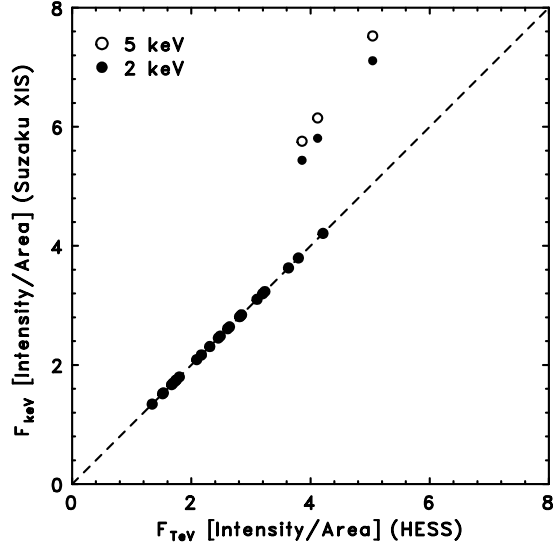}
\caption{The same plot as Figure~\ref{fig:plot_keV_vs_TeV} but results obtained by 
assuming the "toy" model described in the text.}
\label{fig:plot_keV_vs_TeV_toymodel}
\end{figure}

The X-ray flux excess along the NW and SW rims provides a unique probe 
of recent acceleration activity. 
Let us consider a "toy" model and compare its predictions with 
the observational results. 
In the toy model, the injection rate of electrons and protons keeps constant 
but increases by a factor of 1.5 in the last 10~yr only at the NW and SW rims. 
What is important here is the difference of cooling time between electrons and protons. 
The synchrotron cooling time 
of an electron emitting synchrotron photons with energy of $\varepsilon$ is 
given as equation (\ref{eq:sync_cooling_time}), 
while the cooling time of protons due to $p$-$p$ interactions is expressed as 
\begin{eqnarray}
t_{pp} = 5.3 \times 10^7 \left( \frac{n}{1~{\rm cm}^{-3}} \right)^{-1}~{\rm yr}
\end{eqnarray}
and is almost energy-independent. 
For the magnetic field $B=100 \  \rm \mu G$, 
the cooling times of  electrons emitting 2~keV and 5~keV X-rays are
$12~{\rm yr}$ and $7.6~{\rm yr}$, respectively.  
For any reasonable density of  ambient gas, the cooling time of protons is much longer than the 
lifetime of this SNR.  While the synchrotron X-rays we observe at present are emitted by 
electrons accelerated during the last $\sim 10~{\rm yr}$. 
the  flux of  $\pi^0$-decay gamma rays is provided by 
protons accelerated throughout the lifetime of the SNR. 
Figure~\ref{fig:plot_keV_vs_TeV_toymodel} shows 
the scatter plot of $F_{\rm keV}$ and $F_{\rm TeV}$ expected 
from the toy model, which shows a similar distribution to 
the observational results in Figure~\ref{fig:plot_keV_vs_TeV}. 
The recent active acceleration increases the X-ray flux while keeping 
the gamma-ray flux almost unchanged.

\section{SUMMARY}
We observed SNR RX~J1713.7$-$3946 with the {\it Suzaku} observatory. Hard 
X-rays up to $\sim 40~{\rm keV}$ are detected from each of  the 11 pointings. The 
hard X-ray morphology estimated by the HXD PIN  data is generally consistent with an 
extrapolation of that in the energy region below 10~keV. When the HXD spectra 
are fitted with a power law, the photon indices are larger than those obtained from 
the XIS data. 
The difference of photon indices between the XIS and the HXD varies 
from region to region. 
Although this may suggest a variation of spectral shape or
cutoff energy, 
the FoV of the HXD PIN is not small enough to allow detailed imaging 
spectroscopy. Such studies will become 
possible with upcoming missions with hard X-ray mirrors like 
{\it NeXT}, {\it NuSTAR}, or {\it Simbol-X}. 
Moreover these missions  will extend the effective studies up to 100 keV. If the TeV 
gamma-ray emission of RX~J1713.7$-$3946 is of  hadronic origin, this should allow 
detection and study of synchrotron X-ray emission of electrons produced in proton-proton 
interactions via decays of  secondary charged pions \citep{aha_book}. In this regard, the 
above mentioned missions can provide very effective tools for deep (at the level as low as 
$10^{-13} ~{\rm erg}~{\rm cm}^{-2}~{\rm s}^{-1}$) probes of hadronic processes in SNRs in the $\geq 100$ 
TeV energy regime with unprecedented (subarcmin) angular resolution.

Using the XIS and HXD data, we obtained a synchrotron spectrum in the energy 
range of two decades (0.4--40~keV), which means we can probe the parent electron distribution 
in the energy range of one decade regardless of the magnetic field strength. 
The wide-band coverage enables us to see a clear high-energy cutoff in  
the synchrotron spectrum, and for the first time allows detailed studies 
of the cutoff shape and derivation of the spectrum of the parent electrons. 
The spectral shape in the cutoff region was quantitatively 
evaluated, which revealed that the cutoff shape is compatible 
with a theoretical prediction by \cite{zira07}. Based on their model, the cutoff energy 
is obtained as $\varepsilon_0 = 0.67 \pm 0.02~{\rm keV}$. This result, together 
with upper limit of the shock velocity from {\it Chandra} \citep{uchi07}, indicates that 
acceleration efficiency in this remnant approaches the Bohm limit. 

We modeled the multi-wavelength spectrum of RX~J1713.7$-$3946 within both  
the leptonic and hadronic  gamma-ray production scenarios. 
The hadronic model nicely fits the data while it 
seems difficult to explain all observational data with a simple leptonic model. 
The major problem with leptonic models is related to the requirement of 
low magnetic field, $B \leq 15~\mu {\rm G}$ which does not agree 
with the recent discovery of variability of local regions of the shell on year timescales. 
 The hadronic models recently have 
be criticized  based on the lack of thermal emission and low radio flux. 
However, these fact can be accommodated within a standard shock 
acceleration model in a young SNR 
with magnetic field as large as  $100~\mu {\rm G}$, plasma density 
$n \sim 0.1~{\rm cm}^{-3}$,  and distance to the source $d \leq 1~{\rm kpc}$, assuming that
more than 30\% of energy of explosion is released in relativistic protons with 
spectrum as hard as $E^{-2}$ and exponential cutoff around 100~TeV.  
 
 Finally, besides the general strong correlation between X-ray and TeV gamma-ray emission, 
we found an excess X-ray emission compared to gamma-ray emission in the brightest regions of the remnant. 
This excess can be explained by recent activity accompanied by effective acceleration 
of electrons in localised reagions of the shell.  

\acknowledgments
The authors would like to thank all the members of the {\it Suzaku} Science Working 
Group for their help in the spacecraft operation, instrumental calibration, and data 
processing. The authors thank Una Hwang for carefully reading the manuscript 
and Stefan Funk for assisting with the comparison of the morphologies with the H.E.S.S. data 
and for his helpful comments on the manuscript. The authors also thank Misha 
Malkov and Vladimir Zirakashvili for discussions  
related to different aspects of the diffusive shock acceleration theory.   
T. Tanaka and A. Bamba are supported by research fellowships of the 
Japan Society for the Promotion of Science for Young Scientists.

\end{document}